\documentclass[10pt,preprint]{aastex}
\slugcomment{To be Submitted to ApJ}

\begin{document}
%------------------------------------------------------------------------------

\title{ Global General Relativistic MHD Simulation of a Tilted Black-Hole Accretion Disk }

\author{P. Chris Fragile}
\affil{Department of Physics and Astronomy, College of Charleston,
Charleston, SC 29424; fragilep@cofc.edu}
\author{Omer M. Blaes}
\affil{Department of Physics, University of California, Santa
Barbara, CA 93106}
\author{Peter Anninos and Jay D. Salmonson}
\affil{University of California, Lawrence Livermore National
Laboratory, Livermore CA 94550}

%-----------------------------------------------------------------------------

\begin{abstract}
This paper presents a continuation of our efforts to numerically
study accretion disks that are misaligned (tilted) with respect to
the rotation axis of a Kerr black hole. Here we present results of a
global numerical simulation which fully incorporates the effects of
the black hole spacetime as well as magnetorotational turbulence
that is the primary source of angular momentum transport in the
flow. This simulation shows dramatic differences from comparable
simulations of untilted disks. Accretion onto the hole occurs
predominantly through two opposing plunging streams that start from
high latitudes with respect to both the black-hole and disk
midplanes. This is due to the aspherical nature of the gravitational
spacetime around the rotating black hole. These plunging streams
start from a larger radius than would be expected for an untilted
disk. In this regard the tilted black hole effectively acts like an
untilted black hole of lesser spin. Throughout the duration of the
simulation, the main body of the disk remains tilted with respect to
the symmetry plane of the black hole; thus there is no indication of
a Bardeen-Petterson effect in the disk at large. The torque of the
black hole instead principally causes a global precession of the
main disk body. In this simulation the precession has a frequency of
$3 (M_\odot/M)$ Hz, a value consistent with many observed
low-frequency quasi-periodic oscillations. However, this value is
strongly dependent on the size of the disk, so this frequency may be
expected to vary over a large range.
\end{abstract}


\section{Introduction}
\label{sec:intro}

Black-hole accretion has long been postulated to power the energetic
emissions seen from quasars, active galactic nuclei (AGN), and many
galactic X-ray sources; there is now ample observational evidence to
support such claims \citep[e.g. ][]{kro99,mcc05}.  Black-hole
accretion flows are also of interest as laboratories to test
predictions of general relativity. However, the nature of such flows
is complex, involving time-dependent, multi-dimensional dynamics
with generically little symmetry.  Hence numerical simulations play
an integral role in advancing our understanding.

Many simulations of black-hole accretion flows have been carried out
over the past three decades, both in the hydrodynamic
\citep[e.g.][]{wil72,haw84b,haw91} and magnetohydrodynamic (MHD)
\citep[e.g.][]{koi99,gam03a,dev03b} regimes. A common assumption in
nearly all of the work to date has been that the symmetry plane of
the central black hole is aligned with the midplane of the accretion
flow, at least in some averaged sense. However, there is compelling
observational evidence in several black-hole X-ray binaries (BHBs),
e.g. GRO J1655-40 \citep{oro97} and XTE J1550-564
\citep{han01,oro02}, and AGN, e.g. NGC 3079 \citep{kon05}, NGC 1068
\citep{cap06}, and NGC 4258 \citep{cap07}, suggesting that
misaligned (or tilted) black holes may be common \citep[see
also][]{mac02}. This claim relies on the observation of relativistic
bipolar jets (thought to be aligned with the spin axis of the black
hole) that are not perpendicular to the plane of the accretion disk
observed at large scales.

There are also compelling theoretical arguments that many black
holes should be tilted. First, the formation avenues for many
black-hole - disk systems favor, or at least allow for, a tilted
configuration \citep{fra01a}. In stellar mass binaries, the
orientation of the outer disk is fixed by the binary orbit, whereas
the orientation of the black hole is determined by how it became
part of the system, whether through a supernova explosion or
multi-body interaction. If the black hole formed from a member of a
preexisting binary through a supernova, then the black hole could be
tilted if the explosion were asymmetric. If the black hole joined
the binary through multi-body interactions, such as binary capture
or replacement, then there would have been no preexisting symmetry,
so the resulting system would nearly always harbor a tilted black
hole. This same argument can be extended to AGN in which merger
events reorient the central black hole or its fuel supply and result
in repeated tilted configurations.

If an accretion disk is misaligned or tilted, it will be subject to
Lense-Thirring precession.  For an ideal test particle in a slightly
tilted orbit at a radius $r$ around a black hole of mass $M$ and
specific angular momentum $a$, this precession occurs at an angular
frequency $\Omega_{\rm LT}\approx2aM/r^3$.  Close to the black hole,
this is comparable to the orbital angular frequency
$\Omega=(M/r^3)^{1/2}/[1+a(M/r^3)^{1/2}]\approx\Omega_{\rm Kep}$.
However, because of its strong radial dependence, Lense-Thirring
precession becomes much weaker far from the hole. Therefore, a disk
will experience a differential precession that will tend to twist
and warp it.

A warping disturbance can be communicated through a disk in either a
diffusive or wave-like manner. In the diffusive case, the warping is
limited by secular (i.e. ``viscous'') responses within the disk. In
such a case, Lense-Thirring precession is expected to dominate out
to a unique, nearly constant transition radius \citep{bar75,kum85},
inside of which the disk is expected to be flat and aligned with the
black-hole midplane, and outside of which the disk is also expected
to be flat but in a plane determined by the angular momentum vector
of the gas reservoir. This is what we term a ``Bardeen-Petterson''
configuration. Interestingly, data for the two black-hole X-ray
binaries previously mentioned are best fit by disk components with
inclinations that differ from their binary measurements. The
best-fit inclinations are more consistent with inclination
constraints derived from the radio jets \citep{dav06}, possibly
suggesting Bardeen-Petterson configurations. \citet{cap06} also
claim that the observations of NGC 1068 are consistent with the
predictions of the Bardeen-Petterson effect. Confirmation could come
through observations of relativistically broadened reflection
features \citep{fra05c}.

The Bardeen-Petterson result is expected to apply for Keplerian
disks whenever the dimensionless stress parameter $\alpha$
\citep{sha73} is larger than the ratio of the disk semi-thickness
$H$ to the radius $r$ at all radii.  Given that $\alpha$ is usually
considered to be significantly less than one, this implies very
geometrically ``thin'' disks. Unfortunately, current computational
limitations prevent us from conducting global simulations of disks
that are this thin.  On the other hand, the Bardeen-Petterson regime
may not be that common in real disks. Neglecting relativistic
correction factors, the innermost, radiation pressure and electron
scattering dominated portions of radiatively efficient accretion
disks satisfy
\begin{equation}
\frac{H}{r}\sim\varepsilon^{-1}\left(\frac{L}{L_{\rm Edd}}\right)
\left(\frac{r}{r_G}\right)^{-1}, \label{eqhoverr}
\end{equation}
where $\varepsilon\sim0.1$ is the radiative efficiency, $L/L_{\rm
Edd}$ is the luminosity in units of Eddington, and $r_G=GM/c^2$ is
the gravitational radius.  Note that equation (\ref{eqhoverr}) is
independent of whether the stress is chosen to be proportional to
gas pressure, radiation pressure, or some combination of the two. We
therefore conclude that the Bardeen-Petterson regime will be
relevant in radiatively efficient disks near the black hole only for
very small Eddington ratios $L/L_{\rm
Edd}\lesssim\alpha\varepsilon<<1$. Moreover, radiatively less
efficient, geometrically slim and thick flows will clearly not be in
the Bardeen-Petterson regime.

Global simulations of tilted disks that have $H/r>\alpha$ {\em are}
computationally feasible. In this regime Lense-Thirring precession
is expected to produce warps that propagate in a wave-like manner
\citep{pap95a}. In \citet{fra05b} we presented results from the
first fully general relativistic three-dimensional hydrodynamic
numerical studies of tilted thick-disk accretion onto rapidly
rotating (Kerr) black holes. We found that, although Lense-Thirring
precession did cause the disk to warp, the warping only occurred
inside a radius in the disk at which the precession time became
comparable to other dynamical timescales, primarily the azimuthal
sound-crossing time. After the differential warping ended and the
evolution became quasi-static, the disks underwent near solid-body
precession at rates consistent with some low-frequency
quasi-periodic oscillations (QPOs).

In this paper we extend the results of \citet{fra05b} to include
magnetic fields. The inclusion of magnetic fields is important
because it is now widely believed that local stresses within
black-hole accretion disks are generated by turbulence that results
from the magnetorotational instability \citep[MRI; ][]{bal91}. Here
we report on our first global general relativistic MHD (GRMHD)
simulation of a tilted accretion disk around a moderately rapidly
rotating black hole ($a/M=0.9$). The simulation is initialized
starting from the analytic solution for an axisymmetric torus around
a rotating black hole. A weak poloidal magnetic field is added to
the torus to seed the MRI. After the torus is initialized, the black
hole is tilted by an angle $\beta_0=15^\circ$ relative to the disk
through a transformation of the metric. The system is then allowed
to evolve. This paper reports the results as follows: In \S
\ref{sec:methods} we describe the numerical procedures used in this
GRMHD simulation. In \S \ref{sec:results} we present the results of
this simulation. In \S \ref{sec:discussion} we summarize our
findings and draw conclusions.

\section{Numerical Methods}
\label{sec:methods}

This work is carried out using the Cosmos++ astrophysical
magnetohydrodynamics code \citep{ann05}. Similar to our predecessor
code Cosmos \citep{ann03a}, Cosmos++ includes several schemes for
solving the GRMHD equations. The fluid equations can be solved using
a traditional artificial viscosity scheme, non-oscillatory central
difference methods, or a new hybrid dual energy (internal and total)
method. For this work, we use the artificial viscosity formulation,
mainly because of its speed and robustness. With the magnetic fields
we solve the induction equation in an advection-split form and apply
a hyperbolic divergence cleanser to maintain an approximately
divergence-free magnetic field. For clarity and notation sake, we
present the full evolution equations for mass, internal energy,
momentum, and magnetic induction as solved in this work. Throughout
this paper we use units where $G=c=1$ and the metric signature is
($-$,$+$,$+$,$+$). We use the standard notation in which four- and
three-dimensional tensor quantities are represented by Greek and
Latin indices, respectively.

The evolution equations are
\begin{eqnarray}
 \partial_t D + \partial_i (DV^i) &=& 0 ~,  \label{eqn:av_de} \\
 \partial_t E + \partial_i (EV^i) &=&
    - P \partial_t W - \left(P + Q\right) \partial_i (WV^i) ~,
    \label{eqn:av_en} \\
 \partial_t S_j + \partial_i (S_j V^i) &=&
      \frac{1}{4\pi} \partial_t (\sqrt{-g} B_j B^0)
    + \frac{1}{4\pi} \partial_i (\sqrt{-g} B_j B^i)
    \label{eqn:av_mom} \nonumber \\
    & & {} + \left( \frac{S^\mu S^\nu}{2S^0} - \frac{\sqrt{-g}}{8\pi}
             B^\mu B^\nu \right) \partial_j g_{\mu\nu}
    - \sqrt{-g}~\partial_j \left( P + P_B + Q \right) ~, \\
 \partial_t \mathcal{B}^j + \partial_i (\mathcal{B}^j V^i) &=&
    \mathcal{B}^i \partial_i V^j + g^{ij}~\partial_i \psi ~,
      \label{eqn:av_ind} \\
 \partial_t \psi + c_h^2 \partial_i \mathcal{B}^i &=&
 -\frac{c_h^2}{c_p^2} \psi ~, \label{eqn:div_clean}
\end{eqnarray}
where $g_{\mu\nu}$ is the 4-metric, $g$ is the 4-metric determinant,
$W=\sqrt{-g} u^0$ is the relativistic boost factor, $D=W\rho$ is the
generalized fluid density, $V^i=u^i/u^0$ is the transport velocity,
$u^\mu = g^{\mu \nu} u_\nu$ is the fluid 4-velocity, $S_\mu = W(\rho
h + 2P_B) u_\mu$ is the covariant momentum density,
$E=We=W\rho\epsilon$ is the generalized internal energy density, $P$
is the fluid pressure, $Q$ is the artificial viscosity used for
shock capturing, and $c_h$ and $c_p$ are coefficients to determine
the strength of the hyperbolic and parabolic pieces of the
divergence cleanser. There are two representations of the magnetic
field in these equations: $B^\mu$ is the rest frame magnetic
induction used in defining the stress tensor
\begin{equation}
T^{\mu\nu} = \left(\rho h + 2P_B\right) u^\mu u^\nu + \left(P +
P_B\right)g^{\mu\nu} - \frac{1}{4\pi}B^\mu B^\nu
\end{equation}
and
\begin{equation}
\mathcal{B}^\mu = W(B^\mu - B^0 V^\mu)
\end{equation}
is the divergence-free ($\partial \mathcal{B}^i / \partial x^i =
0$), spatial ($\mathcal{B}^0=0$) representation of the field. The
time component of the magnetic field $B^0$ is recovered from the
orthogonality condition $B^\mu u_\mu = 0$
\begin{equation}
B^0 = -\frac{W}{g} \left(g_{0i} \mathcal{B}^i + g_{ij} \mathcal{B}^j
        V^i \right ) ~. \label{eqn:B0}
\end{equation}
The relativistic enthalpy is
\begin{equation}
h = 1 + \frac{\Gamma P}{(\Gamma-1) \rho} + \frac{Q}{\rho}~,
\end{equation}
where we have assumed an equation of state of the form
$P=(\Gamma-1)\rho \epsilon$. Finally, $P_B = \vert\vert B
\vert\vert^2/8\pi = g_{\mu\nu}B^\mu B^\nu/8\pi$ is the magnetic
pressure. We use the scalar $Q$ from \citet{ann05} with $k_q=2.0$
and $k_l=0.3$. We fix the divergence cleanser coefficients to be
$c_h = c_{\rm cfl} \Delta x_{\rm min}/\Delta t$ and $c_p^2 = c_h$,
where $c_{\rm cfl} = 0.7$ is the Courant coefficient, $\Delta x_{\rm
min}$ is the minimum covariant zone length, and $\Delta t$ is the
evolution timestep. For simplicity, we hold the timestep fixed at
$\Delta t = c_{\rm cfl} \Delta x_{\rm min}$ throughout the
simulation.

These GRMHD equations are evolved in a ``tilted'' Kerr-Schild polar
coordinate system $({t},{r},{\vartheta},{\varphi})$. This coordinate
system is related to the usual (untilted) Kerr-Schild coordinates
$({t},{r},{\theta},{\phi})$ through a simple rotation about the
${y}$-axis by an angle $\beta_0$, such that
\begin{equation}
\left( \begin{array}{c} \sin{{\vartheta}}\cos{{\varphi}} \\
                        \sin{{\vartheta}}\sin{{\varphi}} \\
                        \cos{{\vartheta}} \end{array} \right)
 = \left( \begin{array}{ccc} \cos{\beta_0} & 0 & -\sin{\beta_0} \\
                                 0 & 1 & 0 \\
                  \sin{\beta_0} & 0 & \cos{\beta_0} \end{array} \right)
       \left( \begin{array}{c} \sin{{\theta}}\cos{{\phi}} \\
                        \sin{{\theta}}\sin{{\phi}} \\
                        \cos{{\theta}} \end{array} \right) ~.
\label{eqn:tiltarray}
\end{equation}
The full tilted metric terms are provided in \citet{fra05b} [see
also \citet{fra07b}]. The computational advantages of the
``horizon-adapted'' Kerr-Schild form of the Kerr metric were first
described in \citet{pap98} and \citet{fon98b}. The primary advantage
is that, unlike Boyer-Lindquist coordinates, there are no
singularities in the metric terms at the event horizon, so the
computational mesh can extend into the hole's interior. In
principle, this should keep the inner boundary causally disconnected
from the flow, although numerically there is still some
communication.

The simulation is carried out on a spherical polar mesh with nested
resolution layers. The base grid contains $32^3$ mesh zones and
covers the full $4\pi$ steradians. Varying levels of refinement are
added on top of this base layer; each refinement level doubles the
resolution relative to the previous layer. The main simulation,
referenced as Model 915h, has two levels of refinement, thus
achieving a peak resolution equivalent to a $128^3$ simulation. For
comparison we also discuss results from an equivalent untilted
simulation (Model 90h) with the same resolution. As an argument that
our results are reasonably well converged, we also include results
from two other tilted simulations: one with a single refinement
layer and an equivalent resolution of $64^3$ (Model 915m) and
another that starts from a base grid of $24\times24\times32$ and
adds three layers of refinement for an equivalent resolution of
$192\times192\times256$ (Model 915vh). The evolution times for these
simulations differ as described below. In all cases, the full
refinement covers the region $r_{\rm min}\le r \le r_{\rm max}$,
$0.075\pi =\vartheta_1 \le \vartheta \le \vartheta_2 = 0.925\pi$, $0
\le \varphi \le 2\pi$, where $r_{\rm min}=0.98r_{\rm BH}=1.41r_G$
and $r_{\rm max}=120r_G$ are the inner and outer boundaries of the
grid, respectively, and $r_{\rm BH}=1.43r_G$ is the black-hole
horizon radius. The primary motivation for using a nested grid is to
allow us to maintain a reasonable Courant-limited timestep without
sacrificing any spatial resolution within the disk nor completely
excluding the region near the pole. The gain in computational
efficiency is significant since, for a polar mesh, the timestep
scales as $\Delta t \sim r_{\rm min} \sin \vartheta_{\rm min} \Delta
\varphi$. By underresolving the polar region, we gain by increasing
both $\vartheta_{\rm min}$ and $\Delta \varphi$. With 2 levels of
refinement, we are able to use a timestep that is a factor of 11.8
larger than what we could use if our most refined layer extended all
the way to the pole.
%Note, however, that any polar jets that are
%produced in the simulation may not be well resolved.
The main drawback of this approach is that we are unable to resolve
the region in which jets are expected to form.

\begin{figure}
%\begin{center}
%\includegraphics[scale=0.5]{torus3d.m.915h_grid.ps}
%\end{center}
\plotone{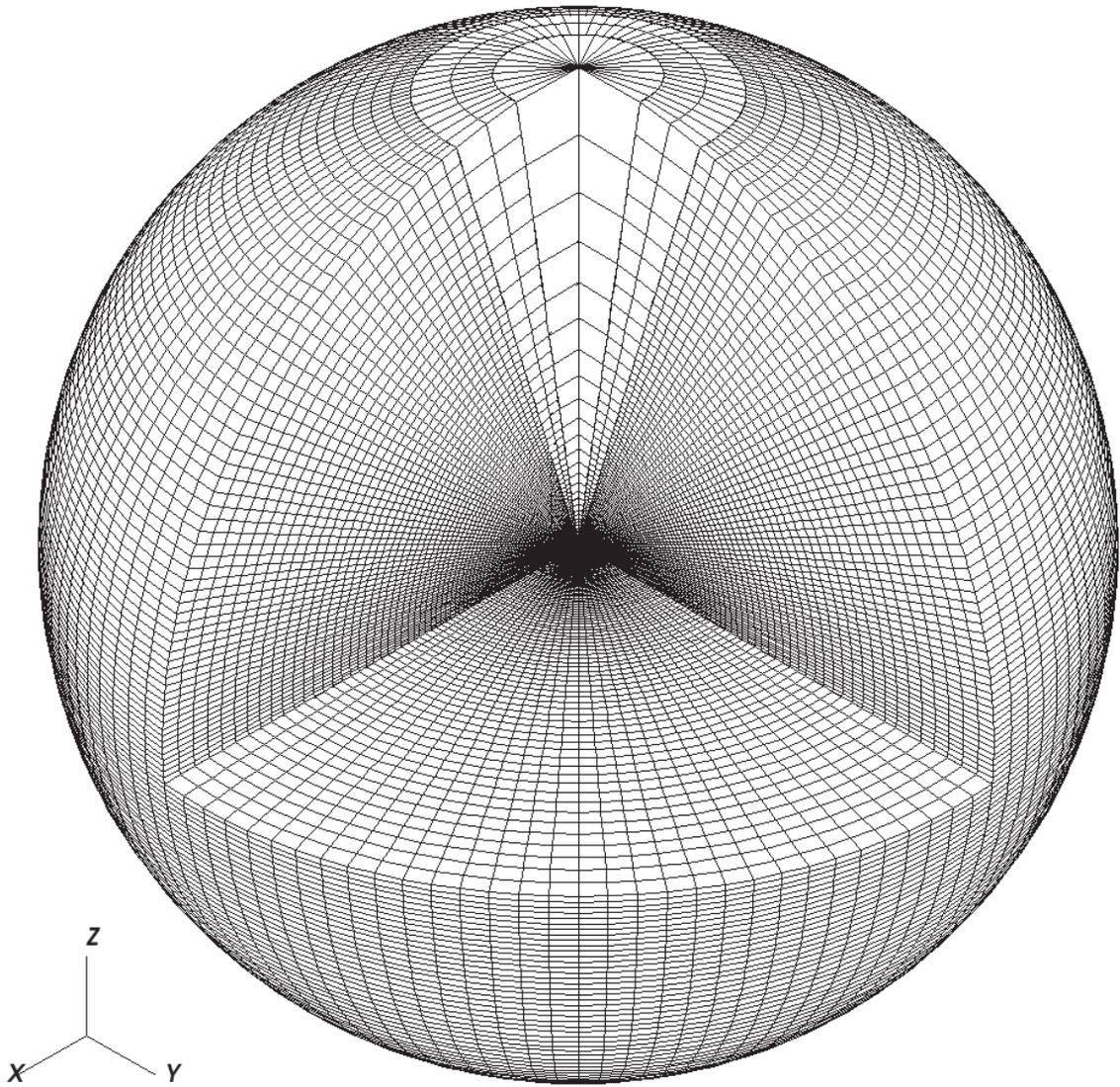} \caption{Plot of the grid geometry used for the
main simulation (Model 915h). The initial torus is aligned in the
symmetry plane of the grid, while the black hole is not.
\label{fig:grid}}
\end{figure}

In the radial direction we use a logarithmic coordinate of the form
$\eta \equiv 1.0 + \ln (r/r_{\rm BH})$. The spatial resolution near
the black-hole horizon is $\Delta r \approx 0.05 r_G$; near the
initial pressure maximum of the torus, the resolution is $\Delta r
\approx 0.5 r_G$. Both are considerably smaller than the initial
characteristic MRI wavelength $\lambda_\mathrm{MRI} \equiv 2\pi
v_\mathrm{A}/\Omega \approx 2.5 r_G$. This also gives us a large
number of zones inside the plunging region. In the angular
direction, in addition to the nested grids, we use a concentrated
latitude coordinate $x_2$ of the form $\vartheta = x_2 + \frac{1}{2}
(1 - h) \sin (2 x_2)$ with $h = 0.5$, which concentrates resolution
toward the midplane of the disk. As a result $r_{\rm center} \Delta
\vartheta = 0.3 r_G$ near the midplane while it is a factor of $\sim
3$ larger for the fully refined zones near the pole. The grid used
in Models 915h and 90h is shown in Figure \ref{fig:grid}.

%\clearpage
%\clearpage

Since we cover the full $4\pi$ steradians, the only ``external''
boundaries are the inner and outer radial boundaries, where we apply
outflow conditions: Fluid variables are set the same in the external
boundary zone as in the neighboring internal zone, except for
velocity, which is chosen to satisfy
\begin{equation}
V^r_\mathrm{ext} = \left\{ \begin{array}{cc}
          V^r_\mathrm{int} & \mathrm{when~} V^r \mathrm{~points~off~the~grid}~, \\
         -V^r_\mathrm{int} & \mathrm{when~} V^r \mathrm{~points~onto~the~grid}~.
         \end{array} \right.
\end{equation}
In the azimuthal direction we apply periodic boundaries at
$\varphi=0$ and $2\pi$. Since Cosmos++ is a zone-centered code, we
do not have to treat the pole ($\vartheta=0$ or $\pi$) directly.
Instead unboosted scalar quantities, such as the gas pressure $P$,
in the ``ghost'' zones across the pole are filled with real data
from the corresponding zone located $180^\circ$ away in azimuth.
Unboosted vector quantities, such as velocity $V^i$, are similarly
filled with data from appropriate real zones, albeit with the signs
reversed for the $\vartheta$ and $\varphi$ components to maintain a
consistent sense of coordinate direction across the pole. Boosted
quantities, since they contain the metric determinant $\sqrt{-g}$,
are reflected across the pole so they extrapolate to zero there.
This treatment differs from the pure reflecting boundaries used in
other works \citep[e.g.][]{dev03c,mck06} in its treatment of the
unboosted variables. For untilted black holes the difference is
relatively minor. However, for tilted black holes, our approach
makes the pole more transparent to the fluid.

We initialize these simulations starting from the analytic solution
for an axisymmetric torus around a rotating black hole
\citep{cha85}. To provide a link with an untilted model already in
the literature, we start with identical torus conditions as model
KDP of \citet{dev03c}, which is the relativistic analog of model GT4
of \citet{haw00}. In our initialization, the torus is defined by:
the black-hole spacetime, specifically the spin of the black hole;
the inner radius of the torus $r_{in}$; the radius of the pressure
maximum of the torus $r_{\rm center}$; and the power-law exponent
$q$ used in defining the specific angular momentum distribution,
\begin{equation}
\ell = -u_\phi/u_t = k \Lambda^{2-q} ~.
\end{equation}
As in model KDP, $a/M=0.9$, $r_{\rm in}=15 r_G$, $r_{\rm center}=25
r_G$, and $q=1.68$. Knowledge of $r_{\rm center}$ leads directly to
a determination of $\ell_{\rm center}$ by setting it equal to the
geodesic value at that radius. The numerical value of $k$ comes
directly from the choice of $q$ and the determination of
$\Lambda_{\rm center}$, where
\begin{equation}
\frac{1}{\Lambda^2} = -\frac{g_{t \phi}+\ell g_{tt}}{\ell g_{\phi
\phi} + \ell^2 g_{t \phi}}~. \label{eqn:Lambda}
\end{equation}
Finally, having chosen $r_{in}$ we can obtain $u_{in}=u_t (r_{in})$,
the surface binding energy of the torus, from $u_t^{-2} =
g^{tt}-2\ell g^{t\phi} +\ell^2 g^{\phi\phi}$.

The solution of the torus variables can now be specified. The
internal energy of the torus is \citep{dev03c}
\begin{equation}
\epsilon(r,\theta) = \frac{1}{\Gamma} \left[ \frac{u_{in}
f(\ell_{in})}{u_t(r,\theta)f(\ell (r,\theta))} \right] ~,
\end{equation}
where $\ell_{in}=\ell(r_{in})$ is the specific angular momentum of
the fluid at the surface and
\begin{equation}
f(\ell) = \left|1-k^{2/n}\ell^\alpha\right|^{1/\alpha} ~,
\end{equation}
where $n=2-q$ and $\alpha=(2n-2)/n$. Assuming an isentropic equation
of state $P=\rho \epsilon(\Gamma-1)=\kappa \rho^\Gamma$, the density
is given by $\rho = \left[ \epsilon(\Gamma-1)/\kappa
\right]^{1/(\Gamma-1)}$. As in model KDP, we take $\Gamma=5/3$ and
$\kappa=0.01$ (arbitrary units). Finally, the angular velocity of
the fluid is specified by
\begin{equation}
\Omega = V^\phi = -\frac{g_{t\phi}+\ell g_{tt}}{g_{\phi \phi} + \ell
g_{t \phi}} ~.
\end{equation}

The dependence of $\Lambda$ on $\ell$ in equation (\ref{eqn:Lambda})
for Kerr black holes means that the solution requires an iterative
procedure. However, we can get an approximate solution by taking the
Schwarzschild form (i.e. ignoring $g_{t \phi}$)
\begin{equation}
\Lambda^2 = -\frac{g_{\phi \phi}}{g_{tt}} ~.
\end{equation}
The error introduced by doing so is small and only affects the
initial torus configuration, which will already be unstable to the
MRI due to the seed magnetic fields being added. Thus, this slightly
simplified treatment has no real consequence for the evolution. We
note that the same procedure is followed in \citet{dev03c}.

Once the torus is constructed, it is seeded with a weak magnetic
field in the form of poloidal loops along the isobaric contours
within the torus. The initial magnetic field vector potential is
\citep{dev03a}
\begin{equation}
A_\varphi = \left\{ \begin{array}{ccc}
          b(\rho-\rho_{\rm cut}) & \mathrm{for} & \rho\ge\rho_{\rm cut}~, \\
          0                  & \mathrm{for} & \rho<\rho_{\rm cut}~.
         \end{array} \right.
\label{eq:torusb}
\end{equation}
The non-zero spatial magnetic field components are then
$\mathcal{B}^r = - \partial_\vartheta A_\varphi$ and
$\mathcal{B}^\vartheta =
\partial_r A_\varphi$. The parameter $\rho_{\rm cut}=0.5*\rho_{\rm max,0}$ is used to keep the field a suitable
distance inside the surface of the torus, where $\rho_{\rm max,0}$
is the initial density maximum within the torus. Using the constant
$b$ in equation (\ref{eq:torusb}), the field is normalized such that
initially $\beta_{\rm mag} =P/P_B \ge \beta_{\rm mag,0}=10$
throughout the torus. This initialization is slightly different than
\citet{dev03b}, who use a volume integrated $\beta_{\rm mag}$ to set
the field strength; the difference is such that $\beta_{\rm
mag,0}=100$ in their work is roughly comparable to $\beta_{\rm
mag,0}=10$ here.

In the background region not specified by the torus solution, we set
up a rarefied non-magnetic plasma accreting into the black hole
\citep{kom06}. The density and pressure have the form
\begin{equation}
\rho = 10^{-3} \rho_{\rm max,0} \exp \left( \frac{-3r}{r_{\rm
center}} \right) ~~,~~ P = \kappa \rho^\Gamma ~.
\end{equation}
The radial velocity has the form
\begin{equation}
V^r = \frac{g^{tr}}{g^{tt}} \left[ 1 - \left( \frac{r_G}{r}
\right)^4 \right] ~.
\end{equation}
This introduces inflow through the horizon without creating large
velocity jumps at the torus surface. This background is initially
more dense than the static background used by \citet{dev03c}.
However, since this background reservoir is not replenished at the
outer boundary, it is rapidly depleted and has virtually no
long-term dynamical impact on the problem. Numerical floors are
placed on $\rho$ and $e$ at approximately $10^{-10}$ and $10^{-16}$
of their initial maxima, respectively. These floors are very seldom
applied once the initial background is replaced by evolved disk
material.

The final step of the initialization is to tilt the black hole by an
angle $\beta_0=15^\circ$ relative to the disk (and the grid) by
transforming the Kerr metric. The full transformation is provided in
\citet{fra05b} [see also \citet{fra07b}]. Thus, while the torus is
responding to the action of the MRI, it will also experience a
gravitomagnetic torque from the tilted black hole.

\section{Results}
\label{sec:results}

\begin{figure}
%\begin{center}
%\includegraphics[scale=0.4]{rho_0.eps}
%\includegraphics[scale=0.4]{rho_1.eps}
%\includegraphics[scale=0.4]{rho_2.eps}
%\includegraphics[scale=0.4]{rho_4.eps}
%\includegraphics[scale=0.4]{rho_7.eps}
%\includegraphics[scale=0.4]{rho_10.eps}
%\end{center}
\plotone{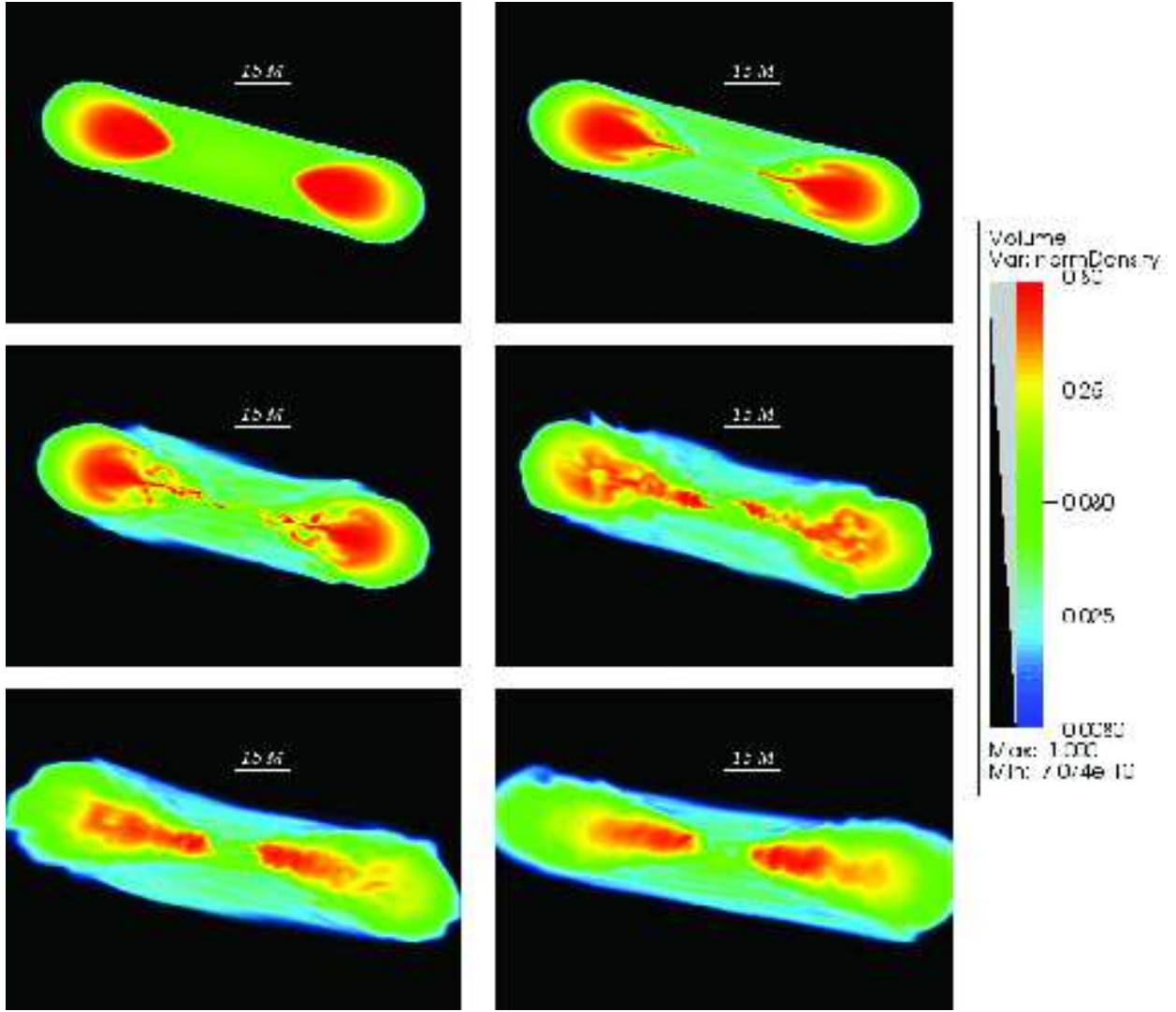}\caption{Volume visualization of the logarithm
of density (scaled from $0.008 \rho_{\rm max,0}$ to $0.8 \rho_{\rm
max,0}$) at ({\em a}) $t=0$, ({\em b}) 1, ({\em c}) 2, ({\em d}) 4,
({\em e}) 7, and ({\em f}) $10t_{\rm orb}$. Half of the disk has
been cut away to reveal the cross section. The black hole spin axis
is oriented vertically in each frame so that the initial torus is
tilted $15^\circ$ to the right. \label{fig:rho_sequence}}
\end{figure}

In the main simulation (915h) the torus is evolved for a total of 10
orbital periods ($10t_{\rm orb}$) as measured at $r=r_{\rm center}$,
which corresponds to $\sim 350$ orbits near $r_{\rm ISCO}=2.32 r_G$,
the coordinate radius of the innermost stable circular orbit (for
prograde orbits in the symmetry plane of the black hole). The very
high resolution simulation (915vh) is only run for half as long ($5
t_{\rm orb}$), while the lower resolution simulation (915m) is run
for twice as long ($20 t_{\rm orb}$). Figure \ref{fig:rho_sequence}
shows snapshots of the disk from Model 915h at times $t=0$, 1, 2, 4,
7, and $10t_{\rm orb}$. The first orbit is dominated by winding of
the magnetic field lines and nonlinear growth of the MRI. Both of
these cause rapid redistributions of disk material and angular
momentum. The initial torus is stretched radially and material
begins to accrete onto the hole and is also carried out to large
radii. A strong current sheet forms in the initial symmetry plane of
the disk through differential winding.

%\clearpage
%\clearpage

From orbits 1-2, MRI driven turbulence begins to grow in the inner
parts of the disk. At the same time, some bending of the disk due to
the differential precession from the hole becomes apparent. The MRI
is fully developed through most of the disk around orbit 2.

By about orbit 7-8, the disk has reached a quasi-steady state. In
the remainder of this section we detail the properties of the
resultant structure. We follow an ``inside-out'' track, starting
from key features of the flow near the hole and working toward
larger radii. Where practical, we draw attention to similarities and
differences between the quasi-steady structure that results in this
simulation and the untilted simulations of \citet{dev03c}. In
particular, we draw attention to the fact that some features, such
as the inner torus and plunging region, are significantly altered,
while others, such as the main body and coronal envelope, show very
similar properties. Again, because of the varying levels of
refinement along the poles, we do not discuss the evacuated funnel
or funnel-wall jet in this paper.

\subsection{Global Structure}

\subsubsection{Plunging Streams}
\label{sec:plunging}

\begin{figure}
%\begin{center}
%\includegraphics[scale=0.75]{rho_10_zoom.eps}
%\end{center}
\plotone{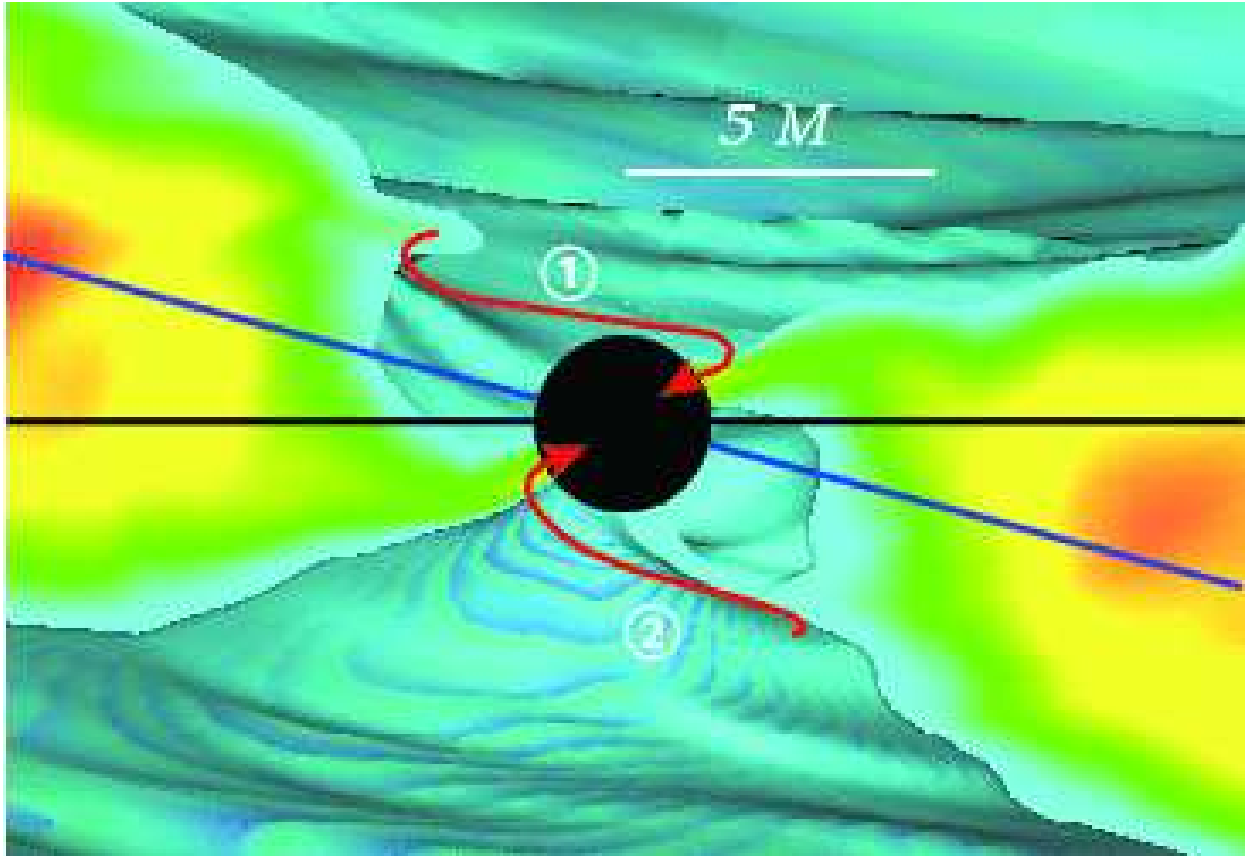}\caption{Zoomed in view of the inner $10 r_G$
of the accretion flow revealing two opposing, high-latitude streams
of material connecting the disk to the horizon (indicated by
arrows). Data is taken from the last frame of the simulation ($t=10
t_{\rm orb}$). To emphasize the plunging streams, the scaling in
this figure is adjusted from that used in Fig.
\ref{fig:rho_sequence} by adding a density isosurface at $\rho =
0.024 \rho_{\rm max,0}$. The figure is oriented as in Fig.
\ref{fig:rho_sequence} with the black-hole spin axis vertical. The
black-hole symmetry plane ({\em black line}) and initial disk
midplane ({\em blue line}) are marked for reference. Note that
stream 1 remains entirely above both planes while stream 2 remains
below. \label{fig:plunge}}
\end{figure}

Perhaps the most striking feature in the tilted disk at late times
are the two opposing streams that start from high latitudes both
with respect to the black-hole symmetry plane and the disk midplane
\citep{fra07a}. Figure \ref{fig:plunge} shows a zoomed-in view of
the region around the black hole including these streams. Note that
stream 1 remains entirely above the black-hole symmetry plane, while
stream 2 remains below. Clearly the material in each stream is in a
plunging orbit into the black hole. Hence, we refer to these
features as the ``plunging streams.''

%\clearpage
%\clearpage

\begin{figure}
%\begin{center}
%\includegraphics[scale=0.5]{torus3d.m.915h_rho_surf.eps}
%\end{center}
\plotone{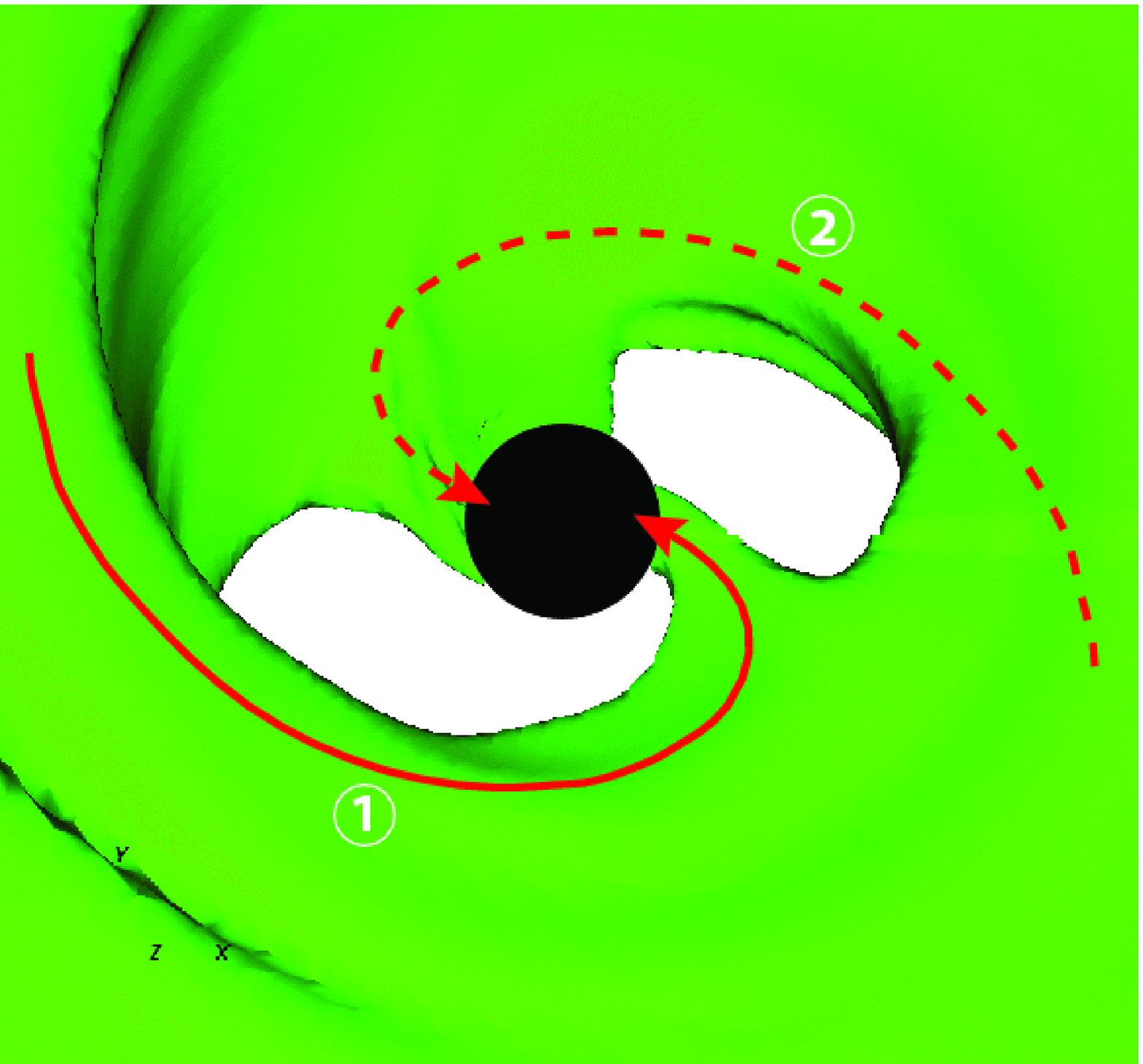}\caption{Isodensity contour at
$\rho=0.1\rho_{\rm max,0}$ from the same time slice as Fig.
\ref{fig:plunge} ($t=10t_{\rm orb}$) viewed down the angular
momentum axis of the black hole. The initial disk angular momentum
axis (and polar axis of the grid) is tilted $15^\circ$ to the right
in this image. One plunging stream (indicated by solid arrow) starts
near the left edge of the figure and connects to the hole on the
right. This stream lies entirely above the black-hole symmetry plane
and corresponds to stream 1 in Fig. \ref{fig:plunge}. The opposing
stream (stream 2) remains below the black-hole symmetry plane and is
seen connecting with the horizon on the left. \label{fig:plunge2}}
\end{figure}

Figure \ref{fig:plunge2} captures the plunging streams from a
different perspective. This image is a view looking down the angular
momentum axis of the black hole onto a single isodensity surface.
The two opposing streams are clearly visible in the interior region
of the disk as well as two relatively evacuated lobes.

%\clearpage
%\clearpage

As material passes through the plunging streams it undergoes strong
differential precession. As we show below, the precession totals
approximately $180^\circ$, accounting for how the material in the
plunging streams is able to enter the black hole from the opposite
azimuth from which it began its plunge without ever passing through
the symmetry plane of the hole.

\begin{figure}
%\begin{center}
%\includegraphics[scale=0.5]{isco2.eps}
%\end{center}
\plotone{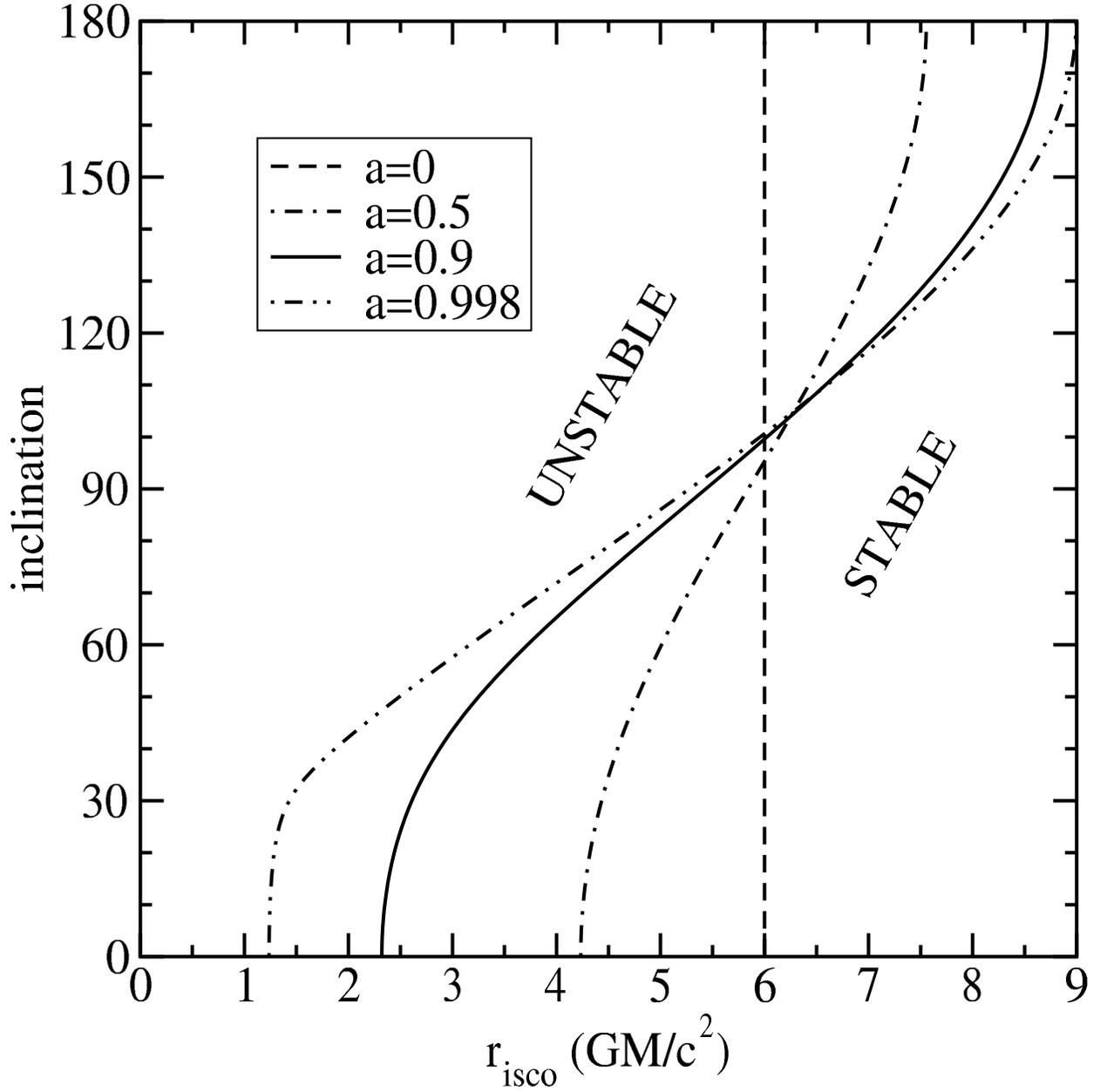}\caption{Plot of the inclination dependence of
$r_{ISCO}$ for black-hole spins $a=0$, 0.5, 0.9, and 0.998.
Inclinations $0 \le i \le 90^\circ$ represent prograde orbits,
whereas inclinations $90^\circ \le i \le 180^\circ$ represent
retrograde orbits. \label{fig:isco}}
\end{figure}

Two very important points to make about these streams is that they
appear to be stable and stationary. They begin forming as early as
$t=7t_{\rm orb}$ and last until the end of the simulation. During
this time their azimuthal location does not change appreciably. The
interesting questions are {\em why} do these opposing plunging
streams form and why do they start from such high latitude with
respect to the black-hole symmetry plane and disk midplane? The
answers, of course, are related and the fundamental cause is the
aspherical nature of the gravitational spacetime around the rotating
black hole. This is best illustrated by considering the dependence
of $r_{\rm ISCO}$ on inclination for orbits that are circular in the
sense that they have constant coordinate radius. Briefly, $r_{\rm
ISCO}$ is the radius at which the quantity
\begin{equation}
R\equiv A^2 \left(\frac{\mathrm{d}r}{\mathrm{d}\tau} \right)^2 =
\left[ E(r^2+a^2)-aL_z \right]^2 - \Delta \left[ r^2 + (L_z-aE)^2 +Q
\right] \label{eq:R}
\end{equation}
and its first two derivatives equal zero, i.e. $R=R'=R''=0$, where
$E$, $L_z$, and $Q$ are the energy, angular momentum, and Carter
constant, respectively, describing orbits around Kerr black holes
\citep{hug01} and $A = r^2 +a^2\cos^2 \theta$ and
$\Delta=r^2-2Mr+a^2$. Following \citet{hug01}, we can eliminate $Q$
in favor of the inclination $i$ defined as
\begin{equation}
\cos i = \frac{L_z}{(L_z+Q)^{1/2}} ~.
\end{equation}
Figure \ref{fig:isco} illustrates this dependence for a few selected
cases of $a$. The key point of the formula and the plot is that
orbital stability around a rotating black hole is strongly dependent
on the inclination of the orbit. Notice that the unstable region
increases monotonically for increasing inclination.

%\clearpage
%\clearpage

\begin{figure}
%\begin{center}
%\includegraphics[scale=0.5]{torus3d.m.915h_rho_slice.eps}
%\end{center}
\plotone{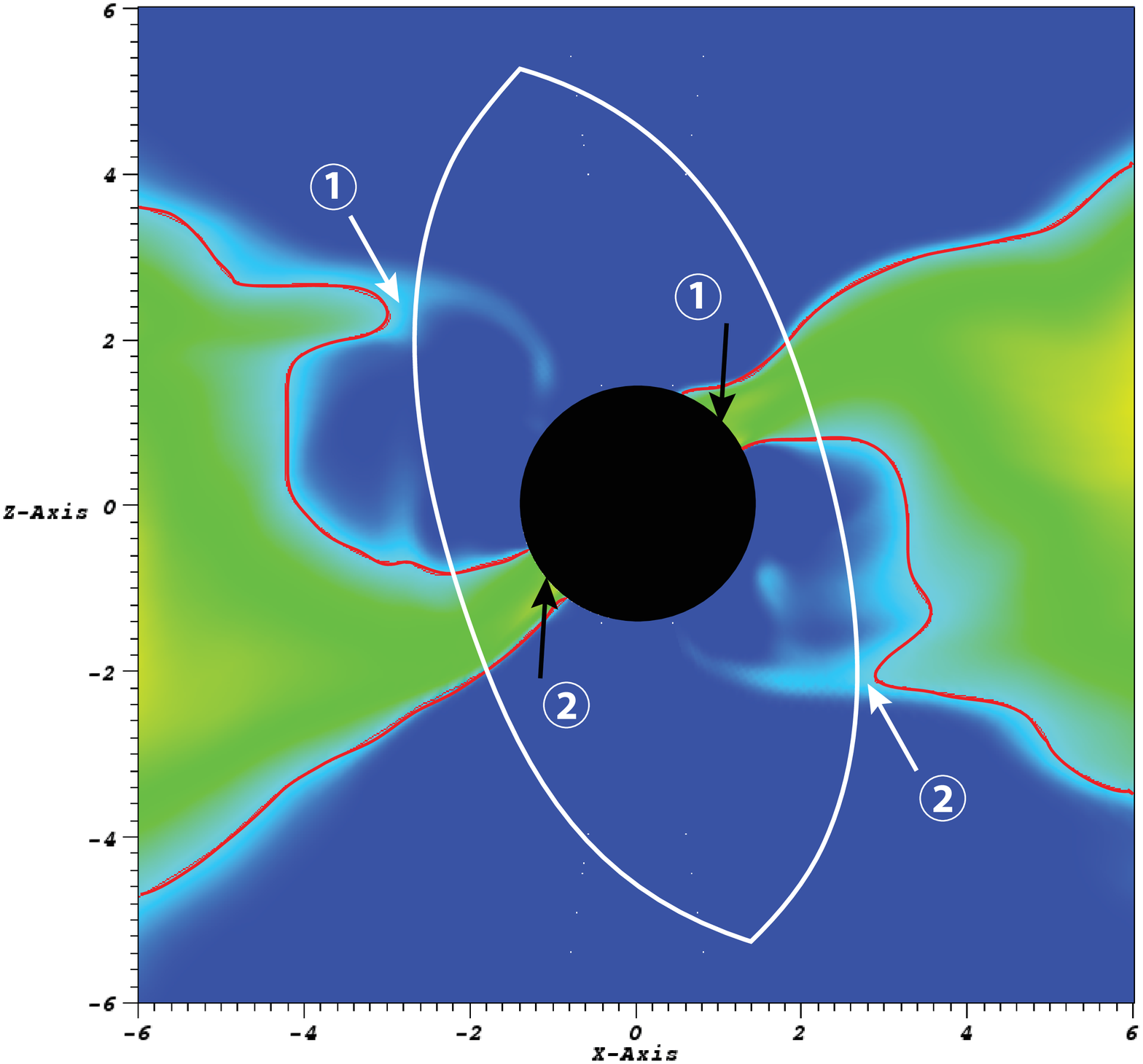}\caption{Meridional plot ($\varphi=0$) through the
final dump ($t=10 t_{\rm orb}$) of the simulation showing a
pseudocolor representation of the logarithm of density (scaled from
$0.008 \rho_{\rm max,0}$ to $0.8 \rho_{\rm max,0}$ as in previous
figures) and an isocontour of density at $\rho = 0.024 \rho_{\rm
max,0}$ (red curve). Unlike previous figures, this one is shown
oriented in the sense of the grid, so that the black hole is tilted
$15^\circ$ to the left. The plot is overlaid with a polar plot of
the ``ISCO surface'' for prograde orbits about an $a=0.9$ black hole
(white curve). This surface is symmetric about the spin axis of the
hole. Notice that the plunging streams from Figs. \ref{fig:plunge}
and \ref{fig:plunge2} start near the largest cylindrical radius
($r\cos\vartheta$) of this surface (indicated by white arrows) and
connect with the horizon approximately $180^\circ$ away in azimuth
(indicated by black arrows). \label{fig:rhoIsco}}
\end{figure}

We can make better use of the information in Figure \ref{fig:isco}
by converting it to a polar plot (using only the prograde orbits)
and overlaying it onto a plot of data from the simulation, as is
done in Figure \ref{fig:rhoIsco}. Such a polar plot creates a
representation of the prograde ``ISCO surface'' (symmetric about the
spin axis of the black hole), which gives a clear indication of
where the most unstable regions of the spacetime are. Note that the
plunging orbits highlighted previously start near where the disk
first encounters the ISCO surface. More precisely, the streams start
near the largest cylindrical radius ($r\cos\vartheta$) of the ISCO
surface, measured with respect to the angular momentum axis of the
disk. This explains why the plunging streams start at such high
inclinations relative to the black-hole symmetry plane and the disk
midplane and why there are only two streams. The plunging region is
no longer azimuthally symmetric from the perspective of the disk.

%\clearpage
%\clearpage

Another point to take away from Figures \ref{fig:isco} and
\ref{fig:rhoIsco} is that $r_{\rm ISCO}$ is larger for larger
inclinations. Thus, for a given black-hole spin, plunging orbits
will always start further away from the hole for more tilted disks.
The tilted black hole effectively acts like an untilted black hole
of lower spin, which would likewise have a larger $r_{\rm ISCO}$.

\subsubsection{Inner Torus}

\begin{figure}
%\begin{center}
%\includegraphics[scale=0.3]{torus3d.m.915h_Rho_av.eps}
%\includegraphics[scale=0.3]{torus3d.m.90h_Rho_av.eps}
%\end{center}
\plottwo{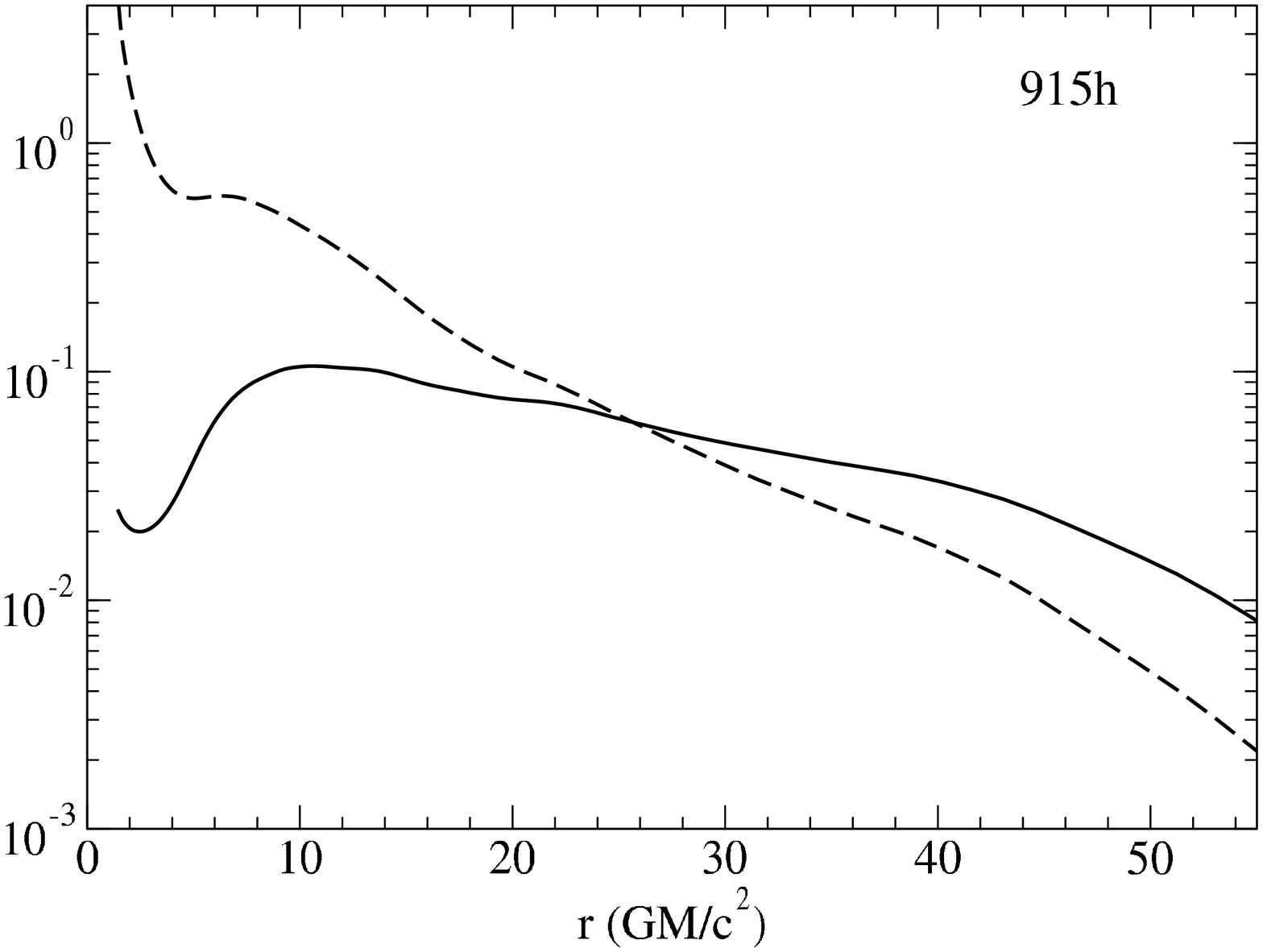}{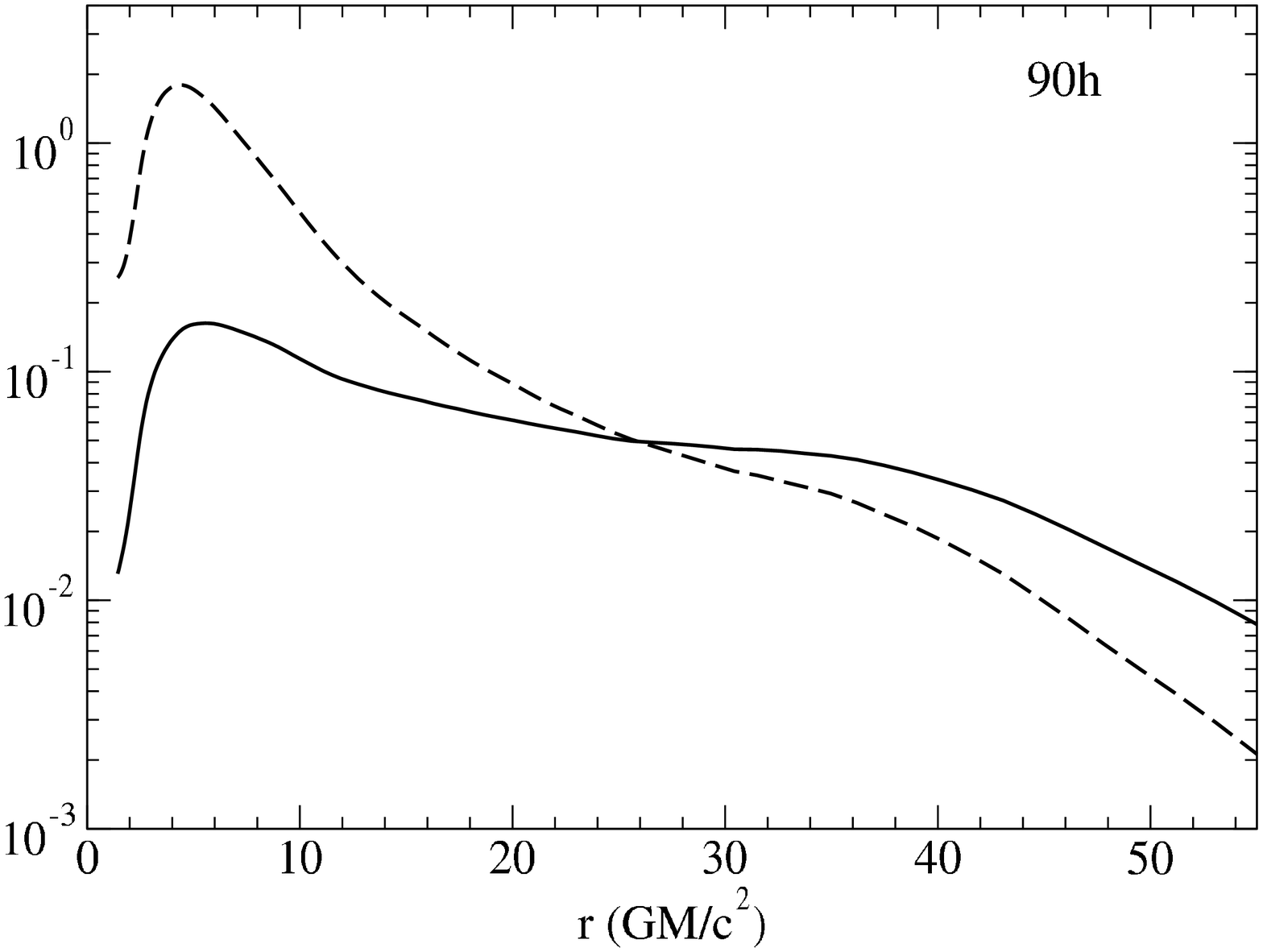}\caption{Plot of
$\langle\langle\rho\rangle_A\rangle_t$ ({\em solid line}) and
$\langle\langle P\rangle_A\rangle_t$ ({\em dashed line}) as a
function of radius for equivalent ({\em a}) tilted
$\beta_0=15^\circ$ (915h) and ({\em b}) untilted $\beta_0=0^\circ$
(90h) simulations. For both simulations, the data has been
time-averaged over the interval $t=9$ to $10t_{\rm orb}$. The
density and pressure have been normalized by their respective maxima
at $t=0$, which are the same in both simulations.
\label{fig:avgdensity}}
\end{figure}

In our tilted simulation, the plunging streams appear to connect
directly to the main disk body without a clearly identifiable
intermediate ``inner torus''. This appears to be a particular result
of the tilted simulation and not, for instance, due to the
differences in the coordinates used in our simulation (Kerr-Schild)
versus those used in \citet{dev03c} (Boyer-Lindquist) or numerical
techniques. We base this statement on the fact that our own untilted
simulation in Kerr-Schild coordinates shows an inner torus very
similar to the one described in \citet{dev03c}. For instance, Figure
\ref{fig:avgdensity} shows the shell-averaged density and pressure
as a function of radius for our tilted and untilted simulations.
Shell averaged quantities are computed over the most refined grid as
follows:
\begin{equation}
\langle\mathcal{Q}\rangle_A(r,t) = \frac{1}{A} \int^{2\pi}_0
\int^{\vartheta_2}_{\vartheta_1} \mathcal{Q} \sqrt{-g}
\mathrm{d}\vartheta \mathrm{d}\varphi ~,
\end{equation}
where $A = \int^{2\pi}_0 \int^{\vartheta_2}_{\vartheta_1} \sqrt{-g}
\mathrm{d}\vartheta \mathrm{d}\varphi$ is the surface area of the
shell. The data in Figure \ref{fig:avgdensity} has also been
time-averaged over the final orbit, $9t_{\rm orb} = t_{\rm min} \le
t \le t_{\rm max} = 10t_{\rm orb}$, where time averages are defined
as
\begin{equation}
\langle\mathcal{Q}\rangle_t = \frac{1}{t_{\rm max} - t_{\rm min}}
\int^{t_{\rm max}}_{t_{\rm min}} \mathcal{Q} \mathrm{d}t ~.
\end{equation}
In the untilted simulation, both the density and the pressure show
local maxima near $4.5 r_G$, indicating an inner torus. The tilted
simulation, on the other hand, shows only marginal evidence for
local maxima near $10 r_G$.

%\clearpage
%\clearpage

\begin{figure}
%\begin{center}
%\includegraphics[scale=0.3]{torus3d.m.915h_Ell_av.eps}
%\includegraphics[scale=0.3]{torus3d.m.90h_Ell_av.eps}
%\end{center}
\plottwo{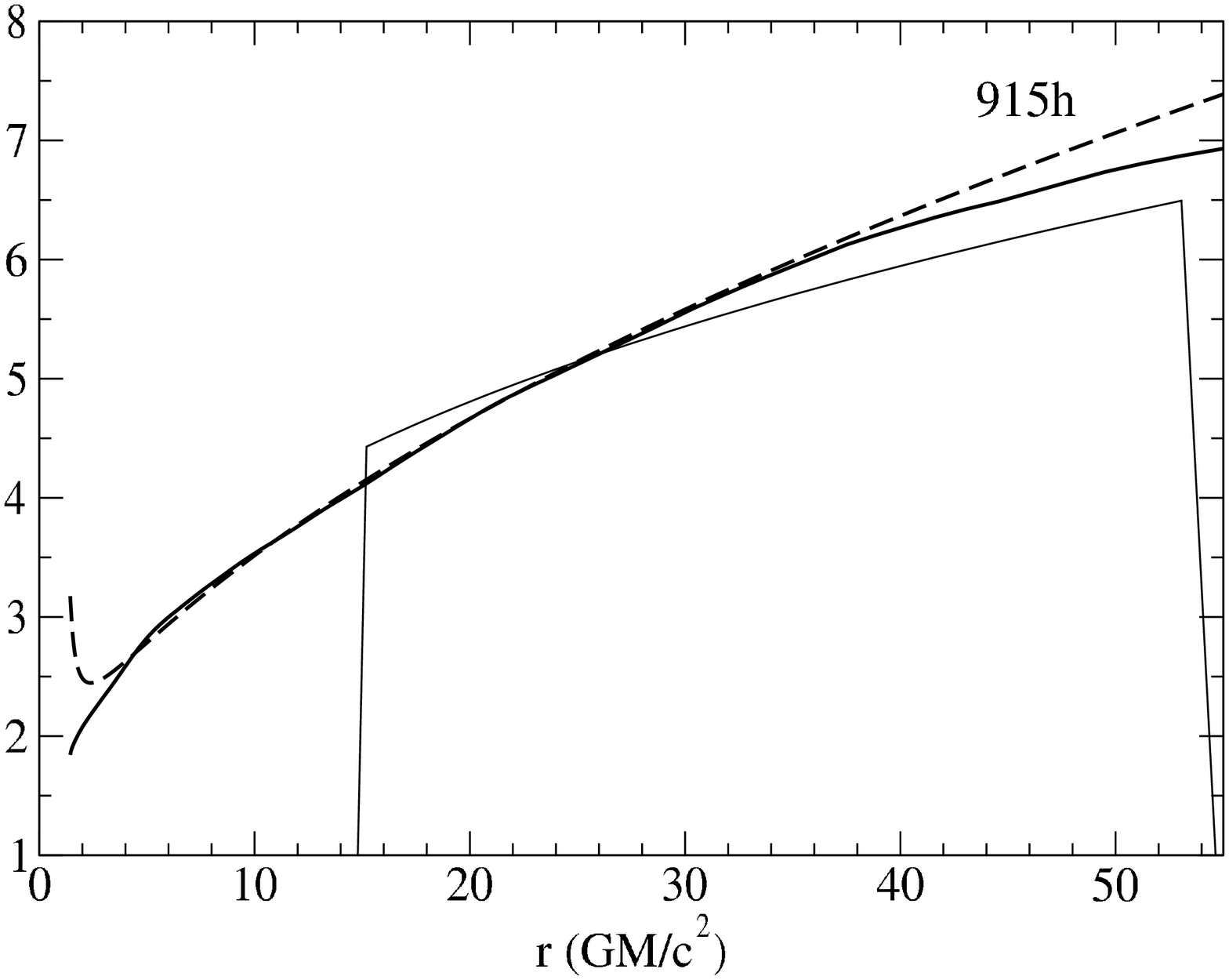}{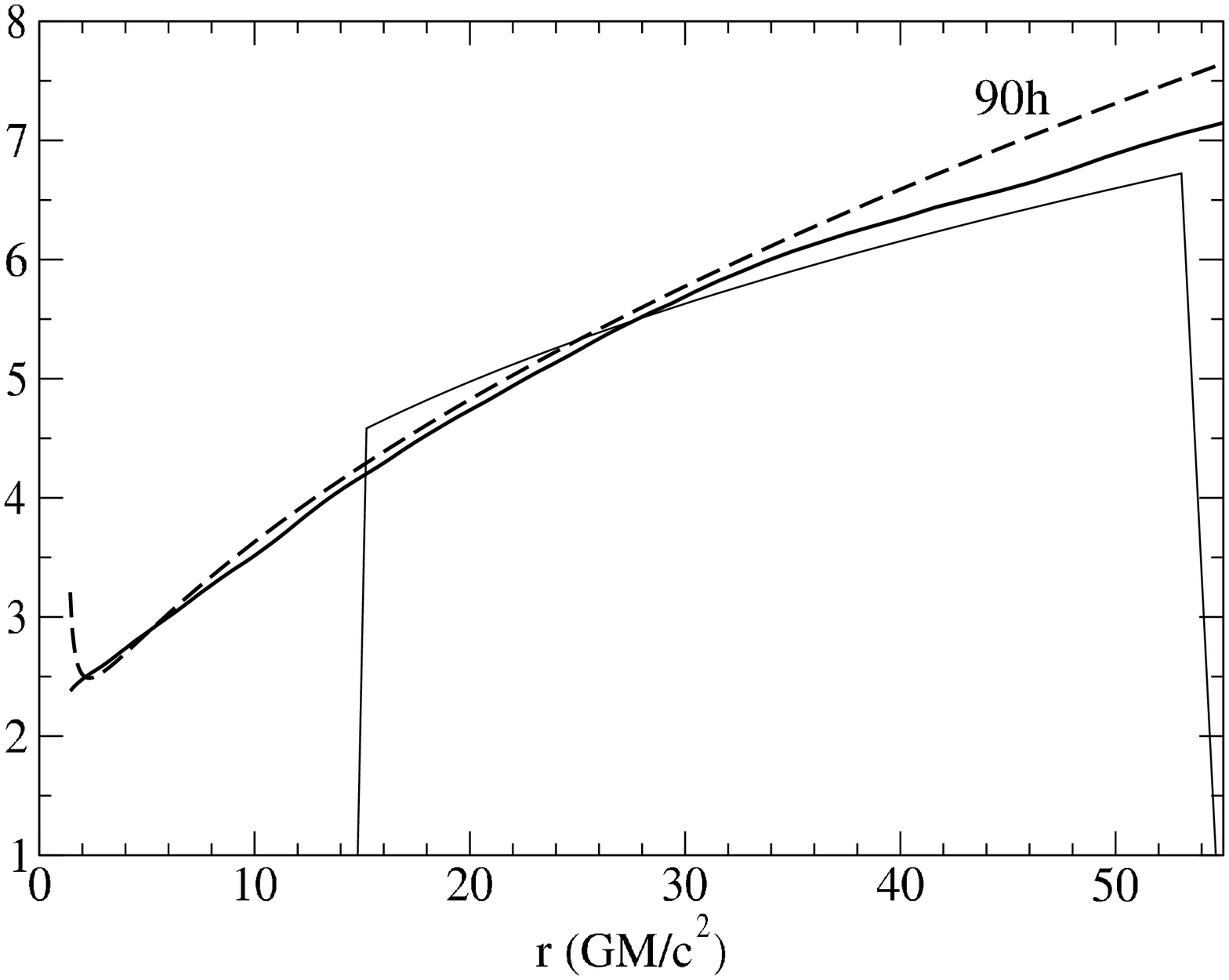}\caption{Plot of the density-weighted
time- and shell-averaged specific angular momentum
$\langle\langle\ell\rangle_A\rangle_t$ ({\em thick line}) as a
function of radius for equivalent ({\em a}) tilted
$\beta_0=15^\circ$ (915h) and ({\em b}) untilted $\beta_0=0^\circ$
(90h) simulations. For both simulations, the data has been
time-averaged over the interval $t=9$ to $10t_{\rm orb}$. In each
plot a comparison is provided with the specific angular momentum of
circular orbits with inclinations of $15^\circ$ and $0^\circ$,
respectively ({\em dashed line}). For reference we also include the
initial angular momentum distribution in the midplane of the torus
({\em thin line}). \label{fig:angmom}}
\end{figure}

Another check of the presence of an inner torus is to look at the
distribution of specific angular momentum in the disk. Because the
inner torus is partially supported by pressure gradients, some
portion of the flow must be locally super-geodesic. In Figure
\ref{fig:angmom} we plot the density-weighted shell average of the
specific angular momentum $\langle\ell\rangle_A = \langle\rho
\ell\rangle_A/\langle\rho\rangle_A$ as a function of radius, again
time-averaged over the interval $t=9$ to $10t_{\rm orb}$. We compare
this against the specific angular momentum distribution of circular
orbits with inclinations of $15^\circ$ and $0^\circ$. These are
calculated from the following expression
\begin{equation}
\ell = \frac{N_1 + \Delta (Mr)^{1/2} N_2^{1/2} \cos i}{D} ~,
\end{equation}
where
\begin{equation}
N_1 = -aMr \left(3r^2 + a^2 - 4Mr \right) \cos^2 i ~,
\end{equation}
\begin{equation}
N_2 = r^4 + a^2 \sin^2 i \left(a^2 + 2r^2 - 4Mr \right) ~,
\end{equation}
and
\begin{equation}
D = a^2 \left( 2r^2 + a^2 - 3Mr \right) \sin^2 i + r^4 + 4M^2r^2 -
4r^3M - Mra^2 ~,
\end{equation}
which comes from noting that for circular orbits $R=R'=0$ from
equation (\ref{eq:R}) and from the definition $\ell=L_z/E$. Both
simulations show a nearly geodesic angular momentum distribution
through most of the disk with a small region of super-geodesic flow
inside $10 r_G$. This region clearly corresponds to the inner torus
in the untilted simulation. It also suggests that there should be an
inner torus in the tilted simulation, though, again, this is not as
evident in the plots of density and pressure.

%\clearpage
%\clearpage

\begin{figure}
%\begin{center}
%\includegraphics[scale=0.6]{Massvst.eps}
%\end{center}
\plotone{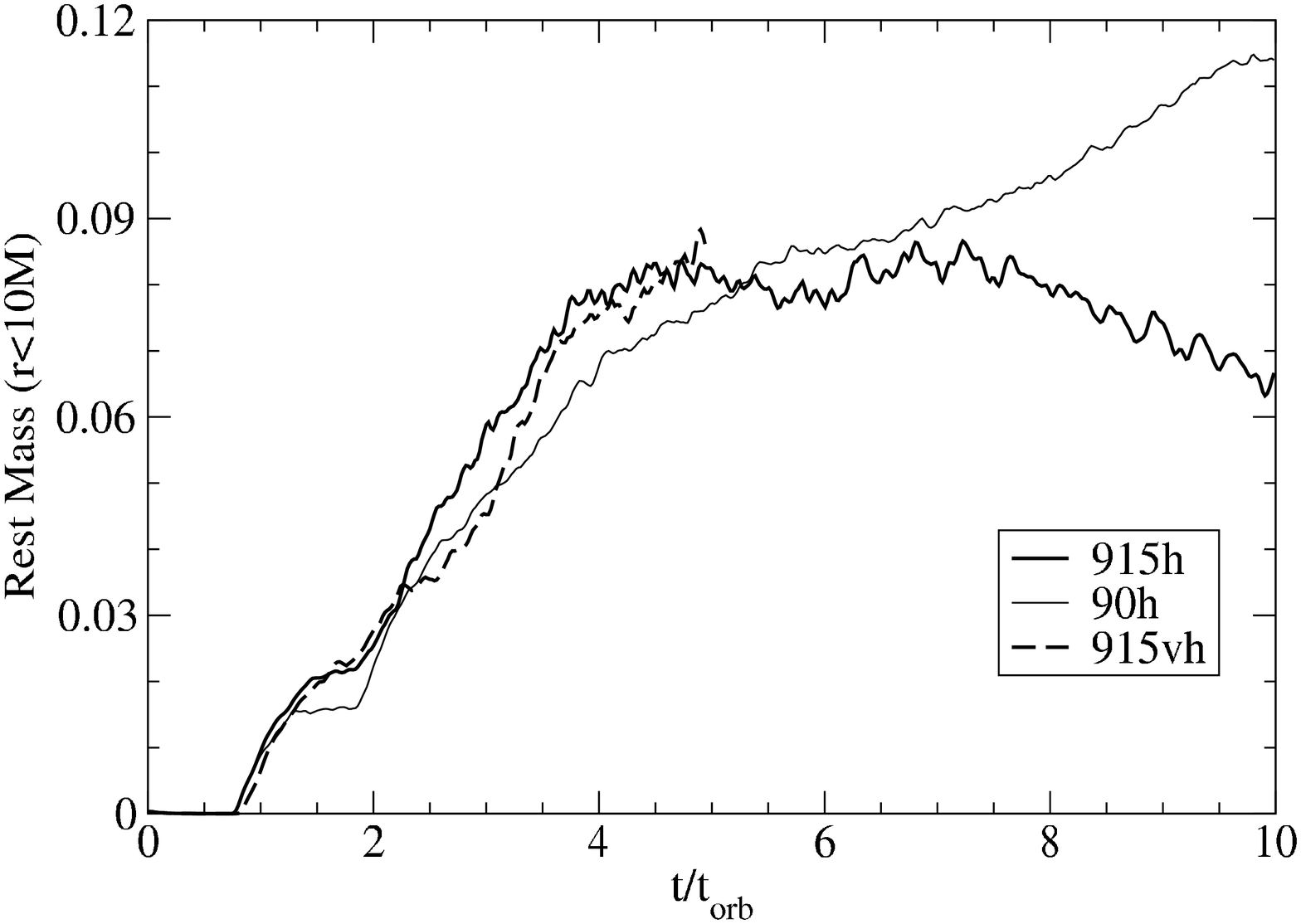}\caption{Total rest mass in the near-hole region
($r<10r_G$) as a function of time for the tilted (915h \& 915vh) and
untilted (90h) simulations. The mass and time are normalized by the
initial mass and orbital period of the torus, respectively.
\label{fig:torusmass}}
\end{figure}

Another indication that the inner torus is less prominent in the
tilted simulation than the untilted one comes from comparing the
total rest mass in the near-hole region ($r<r_{\rm cut} = 10r_G$).
This is done in Figure \ref{fig:torusmass}, where we plot the time
histories of the total (volume-integrated) rest mass
\begin{equation}
\left\langle\rho u^0\right\rangle_V = \int^{2\pi}_0 \int^{\pi}_0
\int^{r_{\rm cut}}_{r_{\rm min}} D \mathrm{d}r \mathrm{d}\vartheta
\mathrm{d}\varphi ~.
\end{equation}
At $t=10t_{\rm orb}$, the inner torus is 42\% less massive in Model
915h.

%\clearpage
%\clearpage

\begin{figure}
%\begin{center}
%\includegraphics[scale=0.3]{massFluxA.eps}
%\includegraphics[scale=0.3]{massFluxB.eps}
%\end{center}
\plottwo{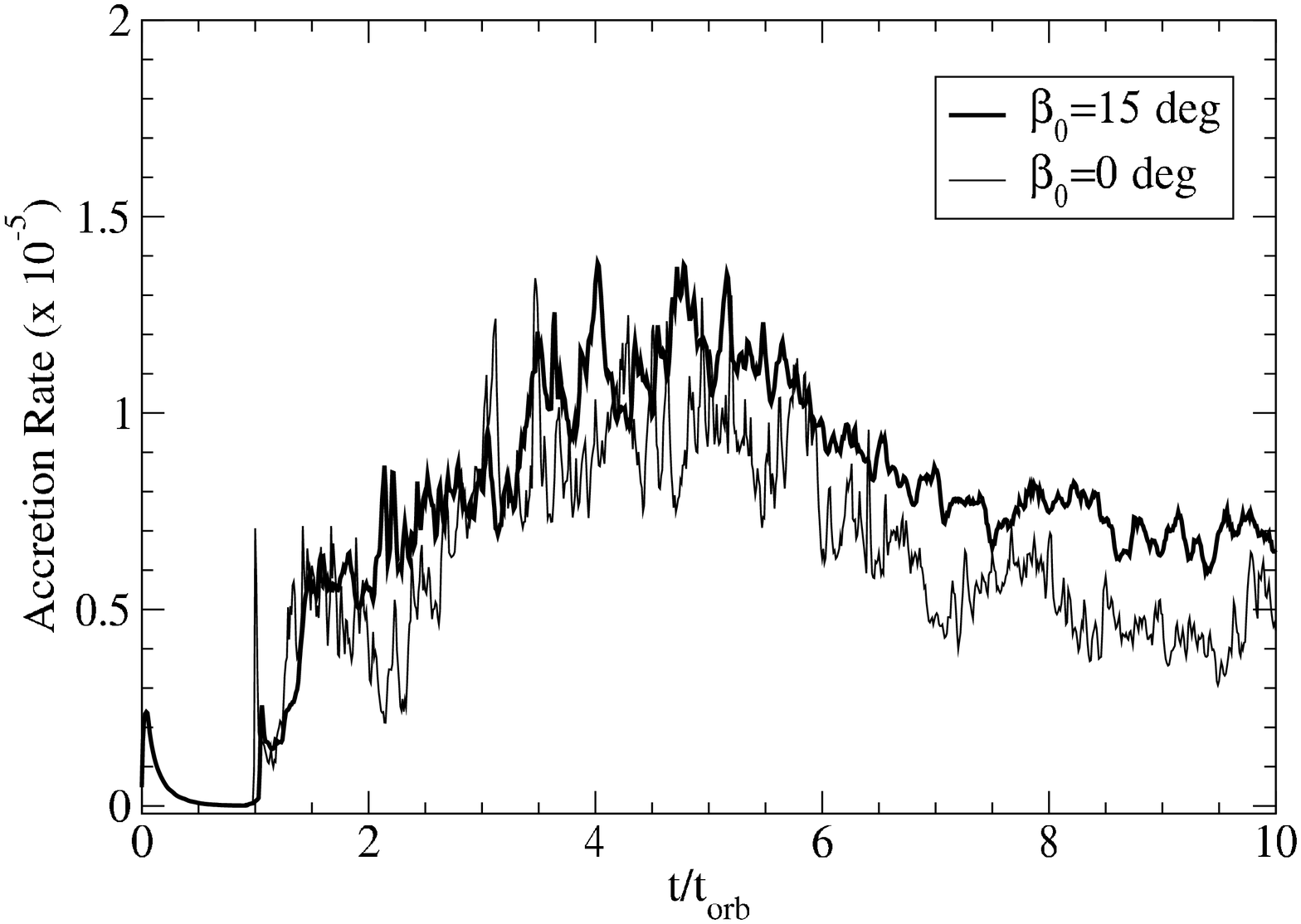}{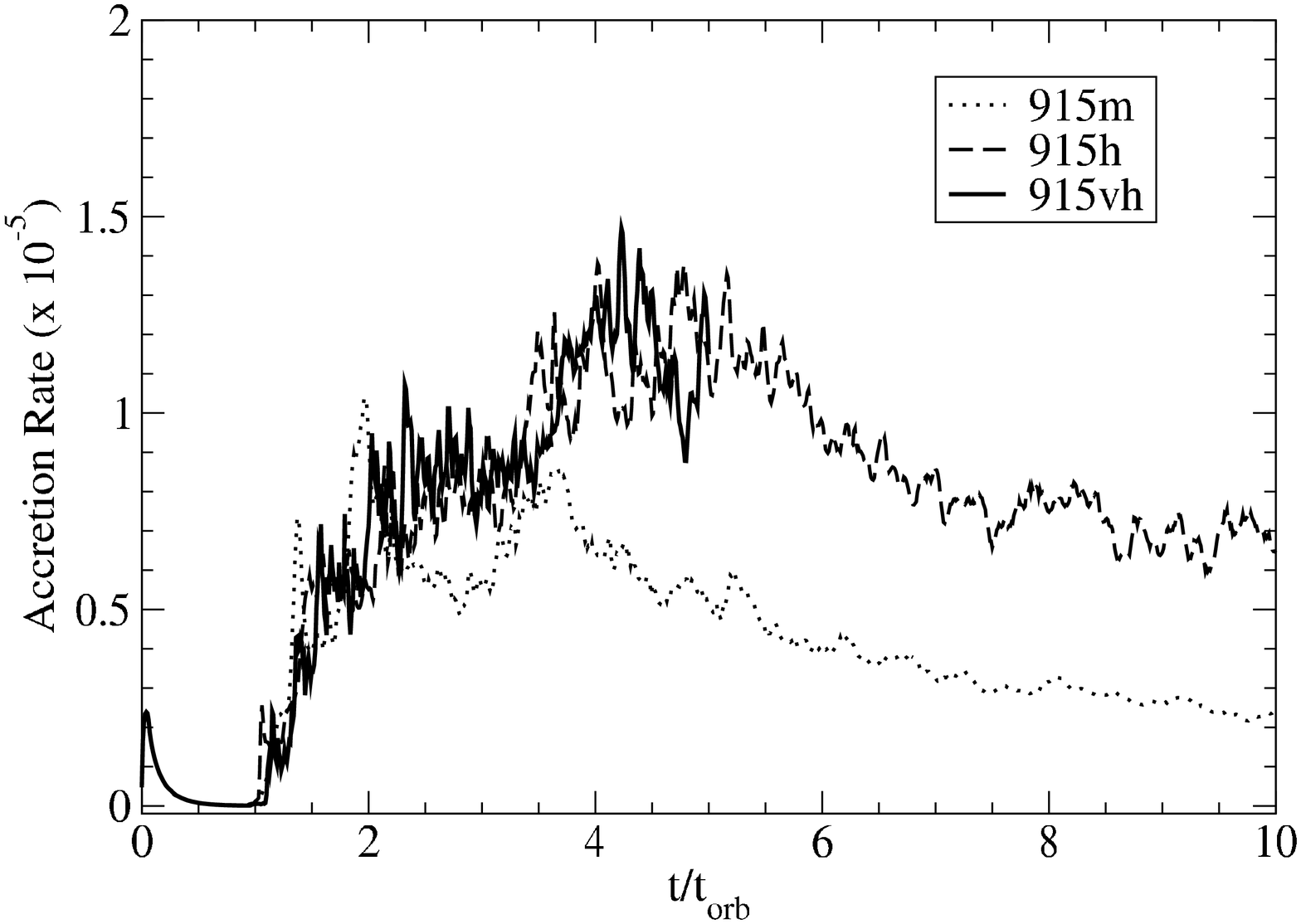}\caption{({\em a}) Plot of the mass
accretion history from Model 915h with $\beta_0=15^\circ$ ({\em
thick line}) and Model 90h with $\beta_0=0^\circ$ ({\em thin line}).
The accretion rate and time are normalized by the initial mass and
orbital period of the torus, respectively. ({\em b}) Plot of mass
accretion rate, comparing our medium (915m), high (915h), and very
high (915vh) resolution tilted disk simulations. The very high
resolution simulation was only run to $t=5t_{\rm orb}$.
\label{fig:accretion}}
\end{figure}

When present, the inner torus usually performs two functions:
regulating the accretion of matter into the black hole and serving
as the launching point for the funnel-wall jet. Therefore, we may
expect a weaker funnel-wall jet (to be discussed in future work) and
a higher mass accretion rate in our tilted-disk simulation relative
to the untilted simulation due to the less prominent inner torus in
the former. We compute the mass accretion rate
\begin{equation}
\dot{M}(r) = \int^{2\pi}_0 \int^\pi_0 D V^r \mathrm{d}\vartheta
\mathrm{d}\varphi
\end{equation}
100 times per $t_{\rm orb}$ (about every $8M$) at each of the
external grid boundaries. Figure \ref{fig:accretion}{\em a} shows a
plot comparing $\dot{M}(r_{\rm min})$ for our equivalent tilted and
untilted simulations. When averaged over the quasi-steady state of
each simulation (from $t=7$ to $10 t_{\rm orb}$),
$\langle\dot{M}\rangle_t$ into the hole for the tilted simulation
(915h) is $7.2\times10^{-6}$, while for the untilted one (90h), it
is $4.9\times10^{-6}$. There is a clear tendancy toward a higher
$\dot{M}$ in the tilted-disk simulation.

%\clearpage
%\clearpage

Figure \ref{fig:accretion}{\em b} compares $\dot{M}$ of the tilted
disk simulation at three different resolutions. Due to the chaotic
nature of the mass accretion we do not expect the individual peaks
to match; yet we are encouraged that the overall shape and magnitude
of the two high-resolution models (915h and 915vh) are very
consistent, suggesting we are reasonably well converged. The medium
resolution simulation (Model 915m), on the other hand, is clearly
underresolved.

\subsubsection{Main Disk Body \& Coronal Envelope}

\begin{figure}
%\begin{center}
%\includegraphics[scale=0.5]{torus3d.m.915h_beta_slice.eps}
%\end{center}
\plotone{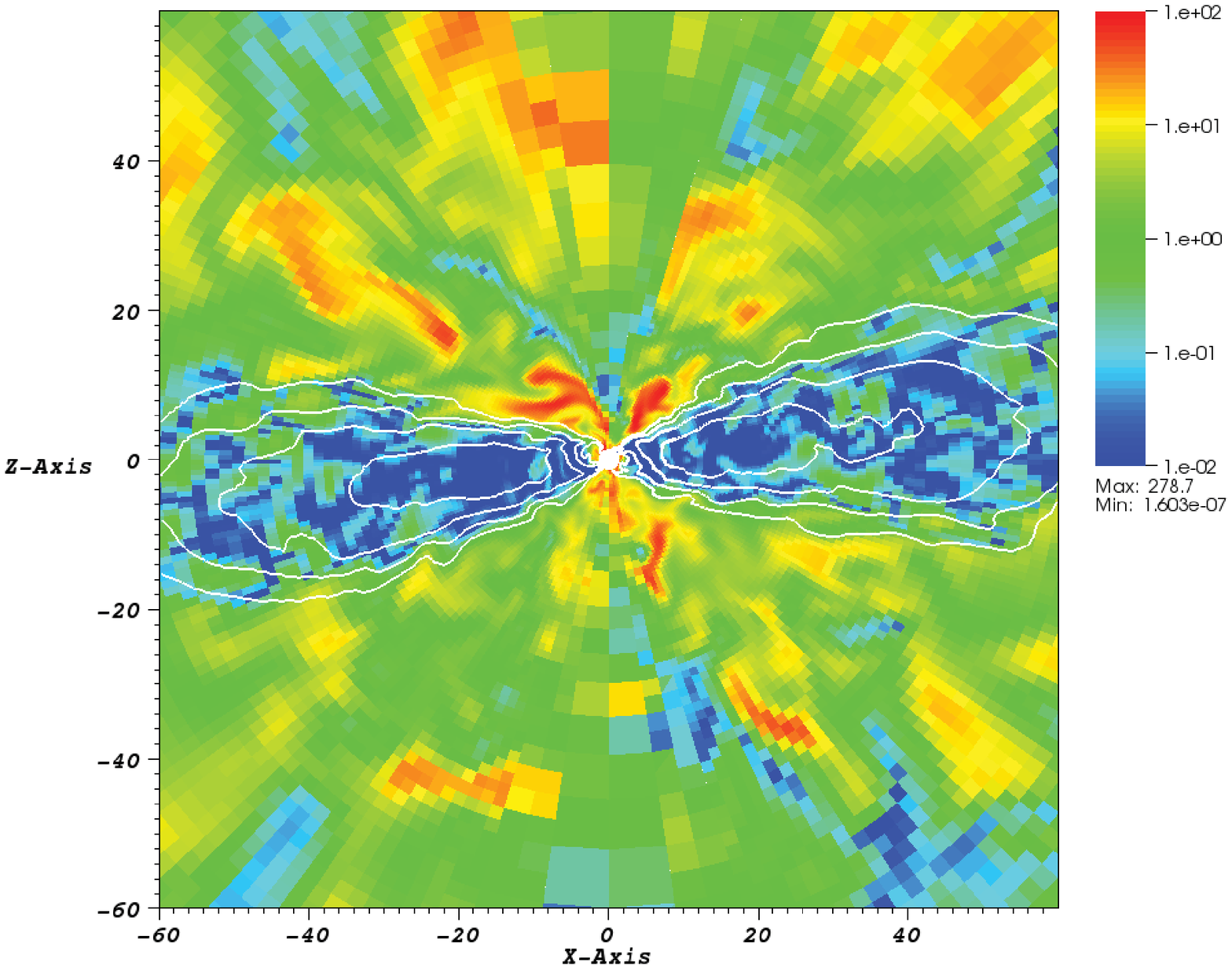}\caption{Azimuthal slice through the simulation
along $\varphi=0$ taken from the final dump ($t=10 t_{\rm orb}$).
The ratio of magnetic pressure to gas pressure ($\beta_{\rm
mag}^{-1}$) is represented as a pseudocolor plot. The colors are
scaled logarithmically and cover the range $10^{-2} \le \beta_{\rm
mag} \le 10^2$. The gas density is given by isocontours at $\rho =
10^{-2}$, $10^{-1.5}$, $10^{-1}$, and $10^{-0.5} \rho_{\rm max,0}$.
As with Fig. \ref{fig:rhoIsco}, this figure is oriented in the sense
of the grid, so that the black hole is tilted $15^\circ$ to the
left. The apparent tilt of the disk is actually due its precession
about the black-hole spin axis, such that the angular momentum axis
of the disk is no longer in the plane of this image; the disk has
not actually realigned with the hole. We remind the reader that the
region near the poles is not sufficiently resolved, so caution
should be used when interpreting results there.
\label{fig:beta_mag}}
\end{figure}

The main disk body does not differ substantially between the tilted
and untilted simulations, except in the notable fact that the tilted
disk precesses (as discussed in \S \ref{sec:precession} below).
Likewise, the coronal envelope, which extends above and below the
disk, shows very similar properties in all our simulations.
% except in the circumpolar region which we describe next.
The material in the coronal envelope is characterized by low density
and rough magnetic equipartition ($\beta_{\rm mag} \approx 1$). By
contrast the main body of the disk is generally gas-pressure
dominated ($\beta_{\rm mag} \ll 1$). Therefore, a plot of
$\beta_{\rm mag}$ and $\rho$, such as Figure \ref{fig:beta_mag},
provides a convenient means to identify these two regions. As found
in \citet{dev03c}, the material in the coronal envelope moves mostly
radially outward, yet has ($-hu_t <1$). This suggests that the
material may be gravitationally bound, in which case it must
circulate back to the disk at large radii. However, we point out
that this definition of binding energy ignores the contribution of
the magnetic fields, so some of this material may in fact escape the
system. We plan to examine outflows from tilted disks more
thoroughly in future work.

%\clearpage
%\clearpage

Because the disk is precessing, its angular momentum axis does not
remain aligned with the grid. Therefore, an azimuthal slice through
the disk at late times, such as Figure \ref{fig:beta_mag}, may give
the impression that the disk has aligned with the symmetry plane of
the black hole when indeed this is not the case. We now turn to the
question of disk alignment and precession.

\subsection{Results Specific to A Tilted Disk}

\subsubsection{Tilt}
One key diagnostic for describing the global response of a tilted
disk subject to Lense-Thirring precession is the tilt between the
angular momenta of the black hole and disk as a function of radius
and time. For example, in the Bardeen-Petterson solution, no time
variability is observed, and the tilt transitions from nearly zero
close to the black hole to a non-zero asymptote at large radii.

As in \citet{fra05b}, we recover the tilt from the simulation data
using the definition
\begin{equation}
\beta(r) = \arccos\left[ \frac{\mathbf{J}_{\rm BH} \cdot
\mathbf{J}_{\rm Disk}(r)} {\vert\mathbf{J}_{\rm BH} \vert
\vert\mathbf{J}_{\rm Disk}(r) \vert} \right] ~,
\end{equation}
where
\begin{equation}
\mathbf{J}_{\rm BH} = \left( -a M \sin\beta_0 \hat{x}, 0, a M
\cos\beta_0 \hat{z} \right)
\end{equation}
is the angular momentum vector of the black hole and
\begin{equation}
\mathbf{J}_{\rm Disk}(r) = \left[ (J_{\rm Disk})_1 \hat{x}, (J_{\rm Disk})_2
\hat{y}, (J_{\rm Disk})_3 \hat{z} \right]
\end{equation}
is the angular momentum vector of the disk in an asymptotically flat
space. This is given by
\begin{equation}
(J_{\rm Disk})_\rho = \frac{\epsilon_{\mu \nu \sigma \rho}
                         L^{\mu \nu} S^\sigma}
                       {2 \sqrt{-S^\alpha S_\alpha}} ~,
\end{equation}
where
\begin{equation}
L^{\mu \nu} = \int \left( x^\mu T^{\nu 0} - x^\nu T^{\mu 0} \right)
\mathrm{d}^3 x ,
\end{equation}
and $S^\sigma = \int T^{\sigma 0} \mathrm{d}^3 x$. The equations for
$L^{\mu \nu}$ and $S^\sigma$ are integrated over concentric radial
shells of the most-refined grid layer, e.g.
\begin{equation}
S^\sigma (r) = \int^{2\pi}_0 \int^{\vartheta_2}_{\vartheta_1}
T^{\sigma 0} \sqrt{-g} \Delta r \mathrm{d}\vartheta
\mathrm{d}\varphi ~.
\end{equation}
The unit vector $\hat{y}$ points along the axis about which the
black hole is initially tilted and $\hat{z}$ points along the
initial angular momentum axis of the disk.

\begin{figure}
%\begin{center}
%\includegraphics[scale=0.35]{BetavsR_av.eps}
%\end{center}
\plotone{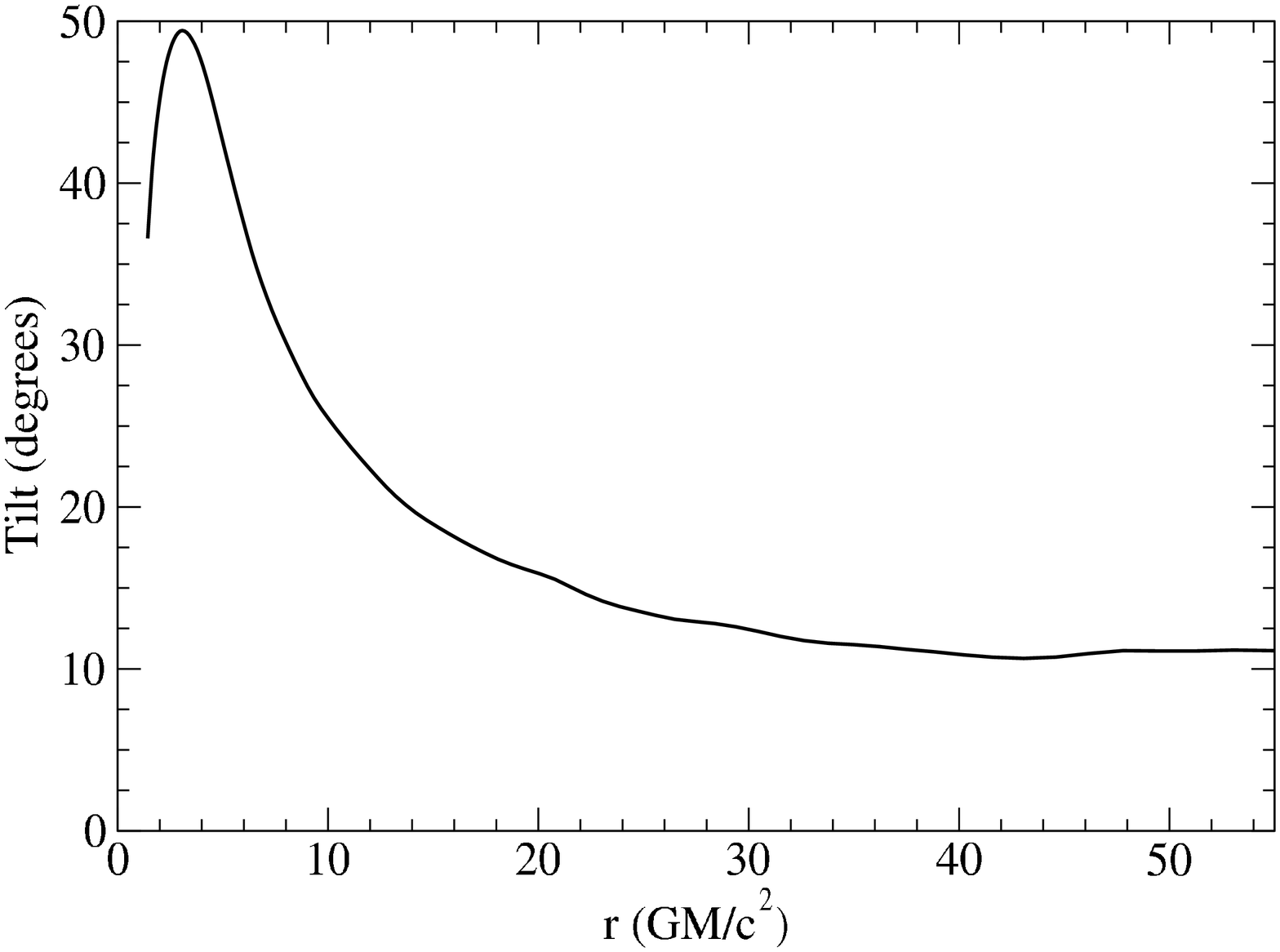}\caption{Plot of the tilt $\langle{\beta}\rangle_t$
as a function of radius through the disk. The data for this plot has
been time averaged from $t=9$ to $10t_{\rm orb}$. The initial tilt
was $\beta_0=15^\circ$. \label{fig:beta}}
\end{figure}

In Figure \ref{fig:beta}, we show the radial profile of $\beta$ time
averaged over the interval $9 t_{\rm orb} \le t \le 10 t_{\rm orb}$.
Recall $\beta_0=15^\circ$ for this simulation. This profile remains
fairly consistent over many orbital times once the quasi-steady
state is reached, so the time-averaged data gives a good
representation for all $t\gtrsim 7 t_{\rm orb}$. The variability
from this time-averaged profile is generally $\lesssim 20\%$ and is
generally carried by moderate amplitude waves traveling through the
disk. The increase in tilt at $r \lesssim 10r_G$ is attributable to
the high latitude plunging streams described in
\S\ref{sec:plunging}.

%\clearpage
%\clearpage

\begin{figure}
%\begin{center}
%\includegraphics[scale=0.35]{scale_height.eps}
%\end{center}
\plotone{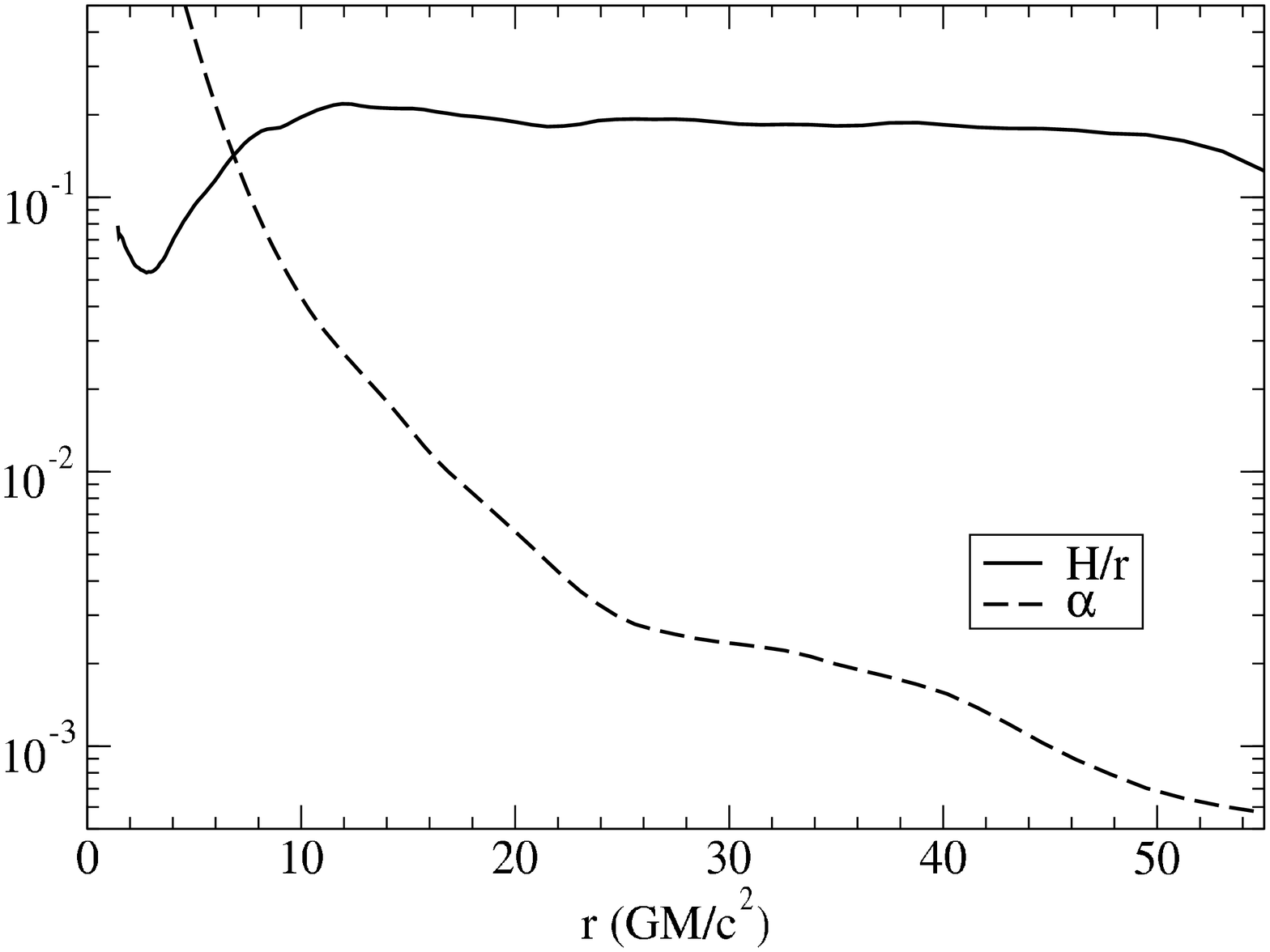}\caption{Plot of the scale height $\langle
H\rangle_t/r$ and magnetic stress parameter
$\langle\alpha\rangle_t$, time averaged over the interval $7 t_{\rm
orb} \le t \le 10 t_{\rm orb}$. This plot shows that this simulation
falls into the thick-disk limit $H/r > \alpha$. \label{fig:stress}}
\end{figure}

One very obvious characteristic of the profile in Figure
\ref{fig:beta} is that $\beta$ does {\em not} approach zero except
perhaps very close to the hole. Thus we do not see evidence for the
Bardeen-Petterson effect in this simulation. This is not surprising
since the Bardeen-Petterson solution is only expected for thin disks
($H/r < \alpha$). This is not the applicable regime for this
simulation, as we illustrate in Figure \ref{fig:stress}, which shows
$H/r$ and $\alpha$ plotted as functions of $r$. The scale height
$H(r)$ is defined in each radial shell as one-half the distance
($0.5 r \Delta \vartheta$) between the two points where $\rho =
\rho_{max}/e$, where we use the time-averaged density along the
$\varphi=0$ azimuthal slice. The dimensionless stress parameter
$\alpha$ in the disk is taken to be
\begin{equation}
\alpha = \left\langle \frac{ \vert u^r u^\varphi \vert\vert B
\vert\vert^2 - B^r B^\varphi \vert}{4 \pi P} \right\rangle_A ~.
\end{equation}
We restrict the calculation of $\alpha$ to only bound material
($-hu_t <1$). Using these definitions we find $H/r \sim 0.2$ and
$\alpha \lesssim 0.01$ through most of the disk.

%\clearpage
%\clearpage

Since warps in slim disks are expected to propagate as bending
waves, it may seem unusual at first that we see little evidence for
such waves in Figure \ref{fig:beta}. For instance, \citet{lub02}
provides an analysis of the theory of bending waves in nearly
Keplerian, weakly inclined disks and predicts that the tilt $\beta$
should be a {\em time-independent, oscillatory} function of radius
\citep[see also][]{mar98}. However, using equation (16) of
\citet{lub02}, we estimate the wavelength of such oscillations for
our simulation to be
\begin{equation}
\lambda \approx \frac{\pi r^{9/4}}{(6a)^{1/2}} \left( \frac{H}{r}
\right) \sim 50 M
\end{equation}
at $r=10 r_G$. This is strongly radially dependent ($\lambda \propto
r^{9/4}$ with $H/r\sim\mathrm{constant}$), so oscillations of
$\beta$ are essentially absent outside $r=10 r_G$, consistent with
what is shown in Figure \ref{fig:beta}.

The same conclusion, that $\beta$ is not expected to oscillate
outside $r=10 r_G$ for this simulation, is also reached by
considering equation (22) of \citet{lub02}. That equation defines a
dimensionless variable
\begin{equation}
x=\left( \frac{24a}{\epsilon^2}\right)^{1/2}
\frac{r^{-(h+1/4)}}{h+1/4} ~,
\end{equation}
which is used to identify the transition radius between oscillatory
behavior and asymptotic behavior, where $h$ and $\epsilon$ are used
to parameterize the radial dependence of the disk scale height
$H/r=\epsilon r^{h-1}$. Whenever $x>>1$ (small $r$), oscillations
should be prominent, whereas whenever $x<<1$ (large $r$), $\beta$
tends to the outer boundary value. For our simulation, with
$\epsilon \approx 0.2$ and $h\approx1$, $x=1$ at $r\approx10 r_G$.
Thus, from both approaches, it is clear that our simulation does not
satisfy the criteria to develop large oscillations in $\beta$ within
the main body of the disk.

Inside $r=10 r_G$, the density of the disk drops off rapidly and the
dynamics are dominated by the plunging streams, which are not
accounted for in the model of \citet{lub02}. Nevertheless, we appear
to capture one-half of one wavelength of a bending wave oscillation
inside $r=10r_G$, based on Figure \ref{fig:beta}. Thus, overall our
results seem to be generally consistent with the predictions of
\citet{lub02}.

\subsubsection{Precession}
\label{sec:precession}

\begin{figure}
%\begin{center}
%\includegraphics[scale=0.35]{GammavsR_av.eps}
%\end{center}
\plotone{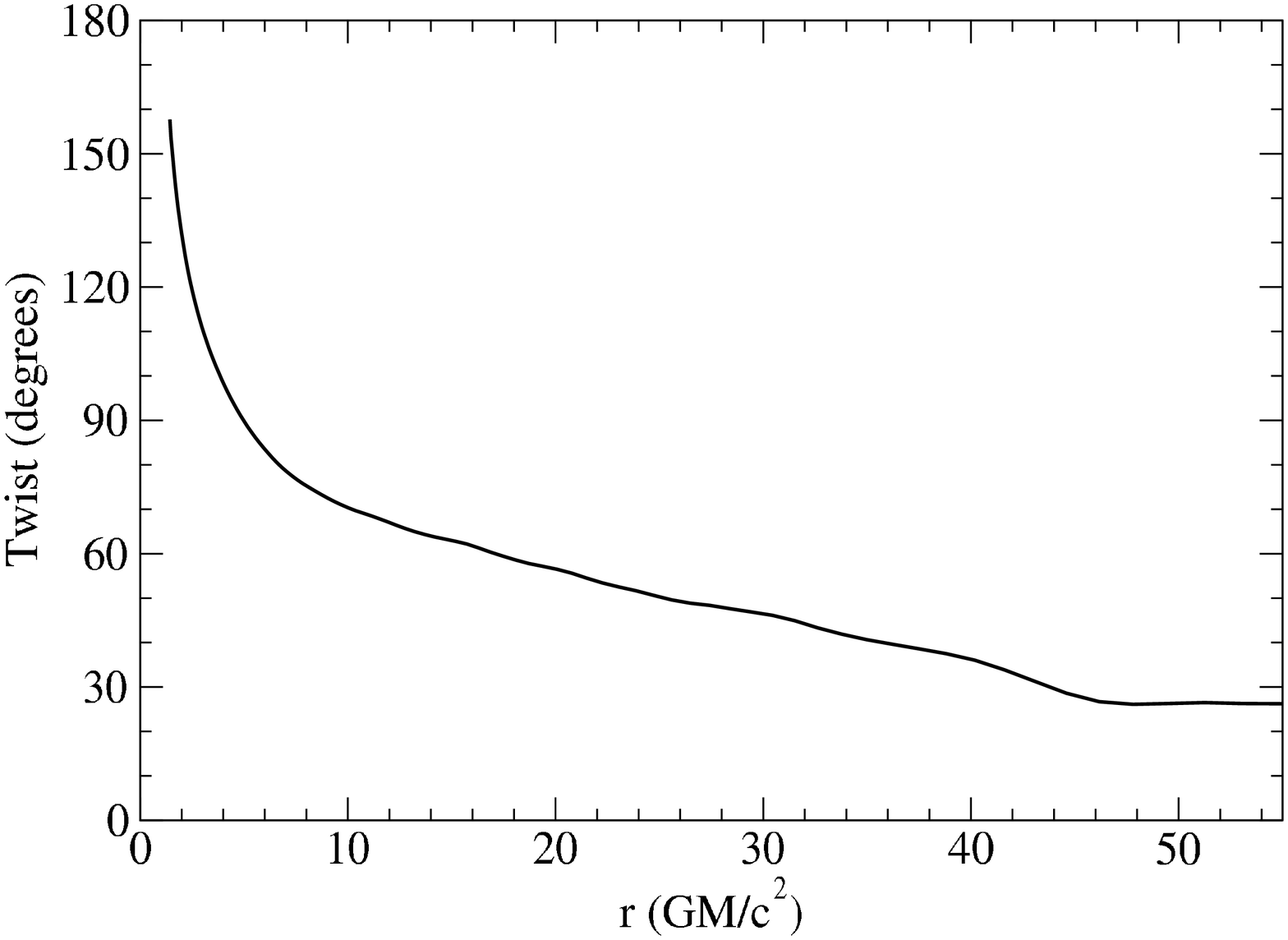}\caption{Plot of the twist $\langle\gamma\rangle_t$
as a function of radius through the disk. The data for this plot has
been time averaged from $t=9$ to $10t_{\rm orb}$. Initially the
twist was zero throughout the disk. The disk matter has precessed
roughly $\sim 180^\circ$ by the time it reaches the hole. The shape
of this twist profile remains fairly constant throughout the
simulation. \label{fig:gamma}}
\end{figure}

A second useful diagnostic for tilted disks is the twist $\gamma$ of
the disk as a function of radius and time. We define the precession
angle (twist) as

\begin{equation}
\gamma(r) = \arccos\left[ \frac{\mathbf{J}_{\rm BH} \times
\mathbf{J}_{\rm Disk}(r)} {\vert \mathbf{J}_{\rm BH} \times
\mathbf{J}_{\rm Disk}(r) \vert} \cdot \hat{y}\right] ~, \label{eq:twist}
\end{equation}
From this definition, $\gamma(r) = 0$ throughout the disk at $t=0$.
In order to capture twists larger than $180^\circ$, we also track
the projection of $\mathbf{J}_{\rm BH} \times \mathbf{J}_{\rm Disk}(r)$ onto
$\hat{x}$, allowing us to break the degeneracy in $\arccos$. A
time-averaged plot of $\gamma$ is provided in Figure
\ref{fig:gamma}.

%\clearpage
%\clearpage

\begin{figure}
%\begin{center}
%\includegraphics[scale=0.35]{timescales.eps}
%\end{center}
\plotone{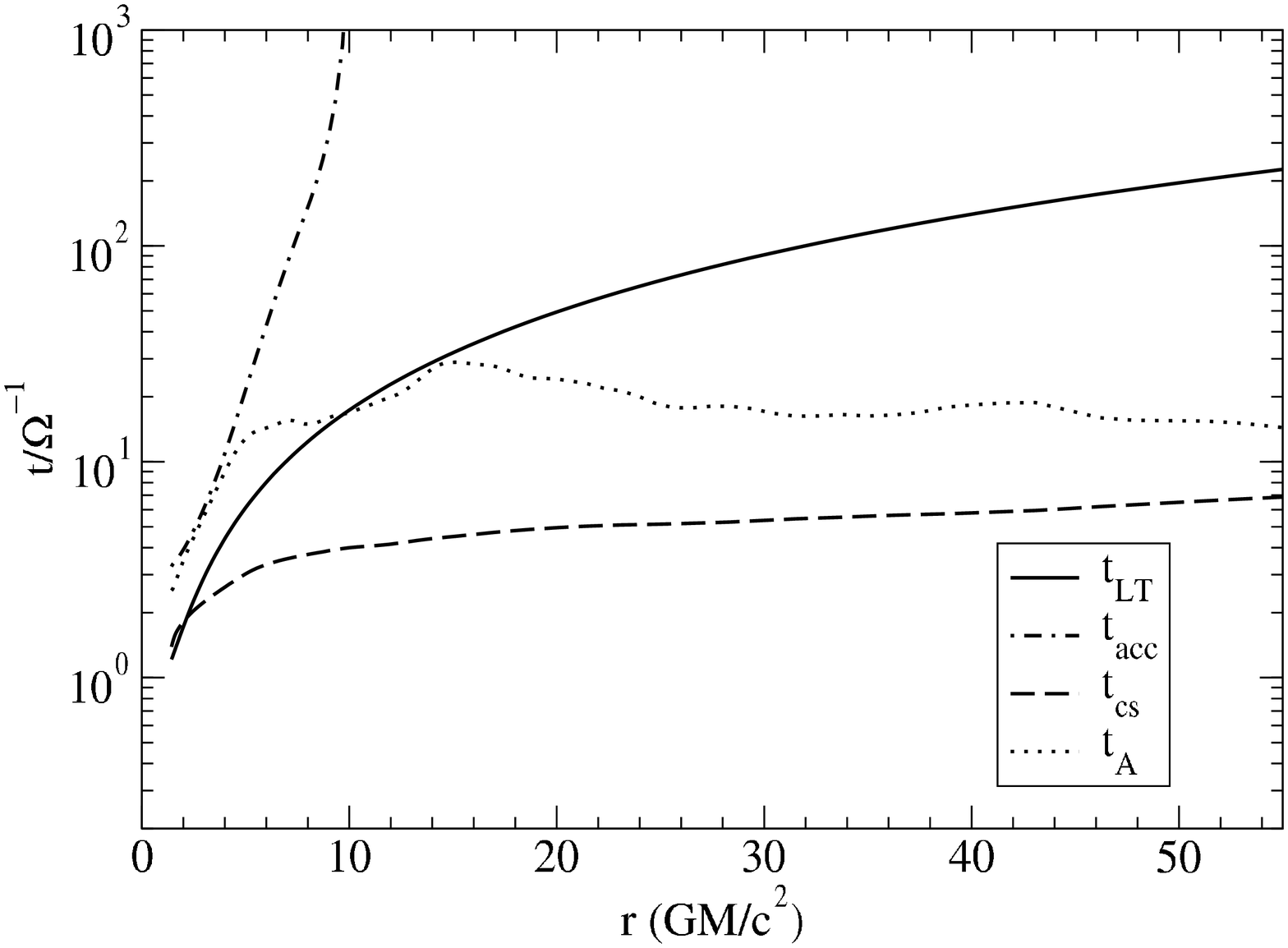}\caption{Plot comparing various timescales within
the disk, including the Lense-Thirring precession timescale $t_{\rm
LT}$, the accretion timescale $t_{\rm acc}$, the sound-crossing time
$t_{\rm cs}$, and the Alfv\'en crossing time $t_{\rm A}$. All
timescales are normalized by the local orbital period in the
midplane of the black hole, $\Omega^{-1}$. The data for this plot
has been time averaged from $t=9t_{\rm orb}$ to $t=10t_{\rm orb}$.
\label{fig:timescales}}
\end{figure}

As described in our previous work \citep{fra05b}, we expect
differential Lense-Thirring precession to dominate whenever the
precession timescale $t_{\rm LT} = \Omega_{\rm LT}^{-1} =
g^{tt}/g^{t\phi}$ is shorter than local dynamical timescales in the
disk \citep{bar75,kum85}. We consider three possible limiting
timescales: the mass accretion timescale $t_{\rm acc} =
r/\overline{V}^r$, where $\overline{V}^r=\langle \langle \rho V^r
\rangle_A/\langle \rho \rangle_A \rangle_t$ is the density-weighted
average inflow velocity; the sound-crossing time
$t_{cs}=r/\overline{c}_s$, where $\overline{c}_s = \langle \langle
\rho c_s \rangle_A/\langle \rho \rangle_A \rangle_t$ is a
density-weighted average of the local sound speed; and the Alfv\'en
crossing time $t_{A}=r/\overline{V}_A$, where $\overline{V}_A$ is a
density-weighted average of the local Alfv\'en speed. The local
sound speed is recovered from the fluid state through the relation
$c_s^2=\Gamma (\Gamma-1)P/[(\Gamma-1)\rho + \Gamma P]$. The Alfv\'en
speed is
\begin{equation}
v_A =  \sqrt{\frac{\vert\vert B \vert\vert^2}{4 \pi \rho h +
\vert\vert B \vert\vert^2}} ~.
\end{equation}
Since $c_s$ and $v_A$ are defined in the frame of the fluid, it is
not strictly accurate to compare $t_{cs}$ and $t_A$ to quantities
defined using the coordinate time (such as $t_{\rm LT}$ and
$\Omega^{-1}$). However, we are mostly concerned with the timescales
in the main body of the disk where such discrepancies are small.
From Figure \ref{fig:timescales}, we can see that the Lense-Thirring
precession timescale is longer than the sound-crossing time at
virtually all radii.

%\clearpage
%\clearpage

\begin{figure}
%\begin{center}
%\includegraphics[scale=0.35]{gammavst.eps}
%\end{center}
\plotone{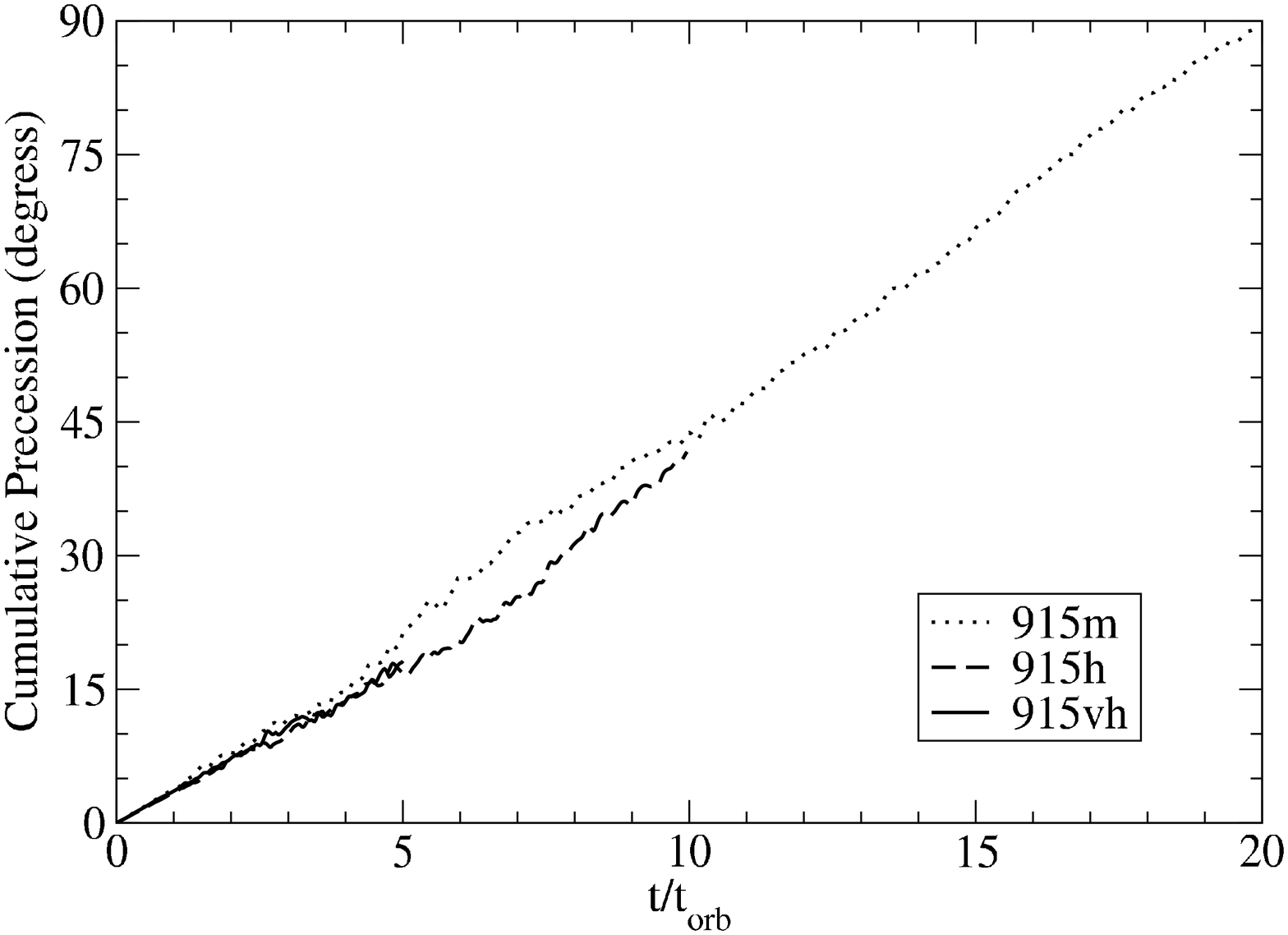}\caption{Plot of the twist $\gamma$, averaged over
the bulk of the disk ($20 \le r/r_G \le 50$), as a function of time.
The slope of this plot can be used to estimate the precession period
of the disk as a whole, which is $0.3 (M/M_\odot)$ s.
\label{fig:precession}}
\end{figure}

Since the sound-crossing time is short compared to the precession
timescale throughout the bulk of the disk, pressure waves strongly
couple the disk material. The disk, thus, responds as a single
entity to the torque of the black-hole and precesses as a global
structure. Such global precession has been noted before in low Mach
number hydrodynamic disks \citep{nel00,fra05b}. To estimate the
precession period, we have plotted $\gamma$, averaged over the bulk
of the disk ($20 \le r/r_G \le 50$), as a function of time in Figure
\ref{fig:precession}. A linear fit to this plot yields a precession
period of $T_{\rm prec} \approx 0.3 (M/M_\odot)$ s, which
corresponds to about $80t_{\rm orb}$. This is longer than the
evolution time of all of our models, so we have had to extrapolate
the full precession period. However, Model 915m is run to $20t_{\rm
orb}$ and shows a nearly linear growth of precession over the full
simulation.

%\clearpage
%\clearpage

Classically, we expect the precession period for a solid-body
rotator with angular momentum $J$ subject to a torque $\tau$ to be
$T_{\rm prec} = 2\pi (\sin \beta) (J/\tau)$ \citep{liu02}. Assuming
a radial dependence to the surface density of the form
$\Sigma=\Sigma_i (r/r_i)^{-\zeta}$ and ignoring higher order general
relativistic corrections, we have $J=2\pi M^{1/2} \Sigma_i r_i^\zeta
r_0^{5/2-\zeta}[1-(r_i/r_o)^{5/2-\zeta}]/(5/2-\zeta)$ and $\tau=4\pi
(\sin \beta) aM^{3/2}
\Sigma_i[1-(r_i/r_o)^{1/2+\zeta}]/[r_i^{1/2}(1/2+\zeta)]$, where
$r_i$ and $r_o$ are the inner and outer radii of the evolved disk,
respectively. Therefore,
\begin{equation}
T_{\rm prec} = \frac{\pi (1+2\zeta)}{(5-2\zeta)}
\frac{r_o^{5/2-\zeta} r_i^{1/2+\zeta} \left[1-(r_i/r_o)^{5/2-\zeta}
\right]} { aM \left[1-(r_i/r_o)^{1/2+\zeta}\right]} ~.
\label{eqn:precession}
\end{equation}
For $r_i=10 r_G$, $r_o=50 r_G$, and $\zeta=0$ (the value we find in
our simulation), equation (\ref{eqn:precession}) predicts $T_{\rm
prec}=0.3 (M/M_\odot)$ s, which is the same as the observed value in
the simulation. Note that equation (\ref{eqn:precession}) differs
from the test particle Lense-Thirring precession period because
$T_{\rm prec}$ depends on the total torque integrated over the
entire disk.

\section{Discussion}
\label{sec:discussion}

In this paper we studied the evolution of an MRI turbulent disk that
was tilted with respect to the spin axis of a modestly fast rotating
black hole. Although this prescription can lead to a
Bardeen-Petterson configuration for some disk parameters, we did not
see evidence for this in this simulation, as alignment of the disk
with the equatorial plane of the black hole did not occur. This is
not surprising since this simulation was carried out in the
thick-disk regime where $H/r > \alpha$ and warps produced in the
disk propagate as waves \citep{pap95a}, rather than diffusively as
in the Bardeen-Petterson case. Since the expected bending wavelength
\citep{lub02} turned out to be longer than the radial extent of the
disk in the simulation, little warping of the disk was observed.
Instead the unwarped disk precessed uniformly. The extrapolated
precession period $T_{\rm prec} \approx 0.3 (M/M_\odot)$ s equates
to periods of $\approx 3$ s and $\approx 3$ d for black holes of
mass $M=10M_\odot$ and $M=10^6 M_\odot$, respectively. Such global
disk precession could explain certain variability features observed
from accreting black holes, such as low-frequency QPOs (LFQPOs)
\citep{ste99,liu02,sch06}, since the observer's viewing angle of the
inner, X-ray emitting region of the disk would vary periodically.
% or the 106 day variability observed from Sgr A* \citep{liu02}.

If the inner disk is optically thick enough to produce
relativistically-broadened reflection features, such as an iron
K$\alpha$ line, then such precession should also be observable
through periodic changes in both the shape and strength of the lines
\citep{fra05c}. These changes should be correlated with the phase of
the corresponding LFQPO. Such a correlation has been observed in GRS
1915+105 \citep{mil05}, although only between line strength and QPO
phase; those data were not sufficiently resolved to determine the
line shape.

Generally, we expect the precession period to be given by equation
(\ref{eqn:precession}), which has a strong dependence on the radial
distribution of the disk ($\propto r_o^{5/2-\zeta}r_i^{1/2+\zeta}$).
One idea to consider is that the outer radius may correspond to the
truncation radius proposed to explain the hard state of black hole
X-ray binaries (e.g. \citet{esi97}, but see also \citet{ryk07}). In
this case our simulated disk would represent the hot, geometrically
thick flow that fills the region inside the truncation radius. The
LFQPO would then correspond to the precession frequency of this
inner flow, in which case it should scale as $r_o^{-5/2+\zeta}$.
\citet{sob00} explored the dependence of the LFQPO frequency on
spectral fitting parameters, including what would be the truncation
radius in the context of the suggested hard state model. They
studied two sources, XTE~J1550-564 and GRO~J1655-40, and found
opposite trends between frequency and radius. For XTE~J1550-564 the
observed frequency was $\nu_{\rm LFQPO}\sim 5$ Hz, and the observed
truncation radius was $r_o/r_G= 2.7 (10M_\odot/M)(D/6
\rm{~kpc})(\cos\theta)^{-1/2}$. From equation (\ref{eqn:precession})
we would expect
\begin{equation}
\frac{r_o}{r_G}=\left[\frac{5-2\zeta}{\pi(1+2\zeta)}\right]^{2/(5-\zeta)}
\left(\frac{a}{M}\right)^{2/(5-\zeta)}
\left(\frac{r_i}{r_G}\right)^{-(1+2\zeta)/(5-2\zeta)}\left(\nu
M\right)^{-2/(5-\zeta)} ~.
\end{equation}
In our simulation we found $\zeta\approx0$, which gives $r_o \approx
33 r_G$ for $M=10 M_\odot$ and $\nu = 5$~Hz. This is considerably
larger than the observed value. However, some of the discrepancy may
be attributable to the large uncertainties in the parameters used to
describe this source, including its distance, mass, and inclination.
Also, if the surface density in XTE J1550-564 depends strongly on
radius, which was not the case for our simulated disk, then our
prediction would change significantly. Further observational studies
along this line are needed to test this prediction more thoroughly.

Although the main body of the disk was not significantly altered by
the tilt, we did find significant differences in the inner regions
of the flow when compared with untilted simulations. First, a tilted
disk encounters the generalized ISCO surface at a larger radius than
an untilted disk. This causes the plunging region to start further
out. The binding energy of the innermost material in the disk is
therefore less than it would be for an aligned disk, and the overall
radiative efficiency should then be reduced.

On the other hand, tilting the disk appears to produce a higher
overall mass accretion rate \citep[shown here in Figure 10a; also
discussed in][]{lod06}. A tilted accretion disk will therefore have
a lower surface density than an untilted disk with the same
accretion rate. This may affect the emergent spectrum, especially
for hot, optically thin flows.  On the other hand for flows that are
effectively optically thick, \citet{dav05} found that the emergent
spectra are remarkably independent of the overall stress and surface
density.

We also found that the plunging region is not axially symmetric.
Instead, accretion onto the hole in the tilted-disk case occurs
through two discrete streams of material that leave the disk at high
latitudes with respect to the black-hole and disk symmetry planes.
%The magnetic linkage between the plunging region and the rest of the
%disk occurs through these discrete high latitude streams;
This may affect the magnitude of magnetic torques exerted by the
plunging region on the disk. An interesting question for future work
is how these streams vary on the timescale of the precession of the
disk. We intend to explore the detailed properties of the plunging
region and innermost disk in a future paper.

The tilted disk also seems not to have formed a clearly identifiable
inner torus. This could be significant because the inner torus
serves as a launching point for the matter-dominated, funnel-wall
jet. The absence of a prominent inner torus may lead to a weaker
matter jet. However, the present simulation is not suited to
addressing this issue because of the poor and varying resolution
used near the pole. Instead, we plan to explore jets and outflows
from tilted disks in future work.

In many respects the tilted disk simulation exhibited properties
consistent with an untilted disk around a black hole of lower spin.
These included the larger plunging radius, higher mass accretion
rate, and less prominent inner torus. Thus black-hole tilt could
hamper efforts to estimate black-hole spin based on such properties.
Indeed, it is commonly stated that astrophysical black hole
spacetimes depend on just two parameters:  mass and spin.  But it
should be remembered that the observed properties of black hole
accretion disks also depend on their inclinations with respect to
the spin axes of their central black holes. This inclination should
be a target of future observational programs that use accretion
disks as surrogates to study properties of black holes.

\acknowledgements We would like to recognize Chris Lindner for his
contributions to this work. We would also like to thank Shane Davis,
Julian Krolik, and the anonymous referee for their suggestions to
improve this manuscript. PCF gratefully acknowledges the support of
a Faculty R\&D grant from the College of Charleston and a REAP grant
from the South Carolina Space Grant Consortium. This work was
supported in part by the National Science Foundation under grants
PHY99-0794 and AST03-07657, under the auspices of the U.S.
Department of Energy by University of California Lawrence Livermore
National Laboratory under contract W-7405-ENG-48, and under the
following NSF programs: Partnerships for Advanced Computational
Infrastructure, Distributed Terascale Facility (DTF) and Terascale
Extensions: Enhancements to the Extensible Terascale Facility.

%\clearpage
\bibliographystyle{apj}
\bibliography{myrefs}

\begin{thebibliography}{}

\bibitem[\protect\citeauthoryear{{Anninos} \& {Fragile}}{{Anninos} \&
  {Fragile}}{2003}]{ann03a}
{Anninos}, P.,  \& {Fragile}, P.~C. 2003, \apjs, 144, 243

\bibitem[\protect\citeauthoryear{{Anninos}, {Fragile}, \&
  {Salmonson}}{{Anninos} et~al.}{2005}]{ann05}
{Anninos}, P., {Fragile}, P.~C.,  \& {Salmonson}, J.~D. 2005, \apj, 635, 723

\bibitem[\protect\citeauthoryear{{Balbus} \& {Hawley}}{{Balbus} \&
  {Hawley}}{1991}]{bal91}
{Balbus}, S.~A.,  \& {Hawley}, J.~F. 1991, \apj, 376, 214

\bibitem[\protect\citeauthoryear{{Bardeen} \& {Petterson}}{{Bardeen} \&
  {Petterson}}{1975}]{bar75}
{Bardeen}, J.~M.,  \& {Petterson}, J.~A. 1975, \apjl, 195, L65

\bibitem[\protect\citeauthoryear{{Caproni} et~al.}{{Caproni}
  et~al.}{2007}]{cap07}
{Caproni}, A., {Abraham}, Z., {Livio}, M.,  \& {Mosquera Cuesta}, H.~J. 2007,
  \mnras

\bibitem[\protect\citeauthoryear{{Caproni}, {Abraham}, \& {Mosquera
  Cuesta}}{{Caproni} et~al.}{2006}]{cap06}
{Caproni}, A., {Abraham}, Z.,  \& {Mosquera Cuesta}, H.~J. 2006, \apj, 638, 120

\bibitem[\protect\citeauthoryear{{Chakrabarti}}{{Chakrabarti}}{1985}]{cha85}
{Chakrabarti}, S.~K. 1985, \apj, 288, 1

\bibitem[\protect\citeauthoryear{{Davis} et~al.}{{Davis} et~al.}{2005}]{dav05}
{Davis}, S.~W., {Blaes}, O.~M., {Hubeny}, I.,  \& {Turner}, N.~J. 2005, \apj,
  621, 372

\bibitem[\protect\citeauthoryear{{Davis}, {Done}, \& {Blaes}}{{Davis}
  et~al.}{2006}]{dav06}
{Davis}, S.~W., {Done}, C.,  \& {Blaes}, O.~M. 2006, \apj, 647, 525

\bibitem[\protect\citeauthoryear{{De Villiers} \& {Hawley}}{{De Villiers} \&
  {Hawley}}{2003a}]{dev03a}
{De Villiers}, J.,  \& {Hawley}, J.~F. 2003a, \apj, 589, 458

\bibitem[\protect\citeauthoryear{{De Villiers} \& {Hawley}}{{De Villiers} \&
  {Hawley}}{2003b}]{dev03b}
{De Villiers}, J.,  \& {Hawley}, J.~F. 2003b, \apj, 592, 1060

\bibitem[\protect\citeauthoryear{{De Villiers}, {Hawley}, \& {Krolik}}{{De
  Villiers} et~al.}{2003}]{dev03c}
{De Villiers}, J., {Hawley}, J.~F.,  \& {Krolik}, J.~H. 2003, \apj, 599, 1238

\bibitem[\protect\citeauthoryear{{Esin}, {McClintock}, \& {Narayan}}{{Esin}
  et~al.}{1997}]{esi97}
{Esin}, A.~A., {McClintock}, J.~E.,  \& {Narayan}, R. 1997, \apj, 489, 865

\bibitem[\protect\citeauthoryear{{Font}, {Ib{\' a}{\~ n}ez}, \&
  {Papadopoulos}}{{Font} et~al.}{1998}]{fon98b}
{Font}, J.~A., {Ib{\' a}{\~ n}ez}, J.~M.~.,  \& {Papadopoulos}, P. 1998, \apjl,
  507, L67

\bibitem[\protect\citeauthoryear{{Fragile} \& {Anninos}}{{Fragile} \&
  {Anninos}}{2005}]{fra05b}
{Fragile}, P.~C.,  \& {Anninos}, P. 2005, \apj, 623, 347

\bibitem[\protect\citeauthoryear{{Fragile} \& {Anninos}}{{Fragile} \&
  {Anninos}}{2007}]{fra07b}
{Fragile}, P.~C.,  \& {Anninos}, P. 2007, \apj, 666, xxx

\bibitem[\protect\citeauthoryear{{Fragile} et~al.}{{Fragile}
  et~al.}{2007}]{fra07a}
{Fragile}, P.~C., {Anninos}, P., {Blaes}, O.~M.,  \& {Salmonson}, J.~D. 2007,
  in proceedings of the 11th Marcel Grossmann Meeting on General Relativity
  (astro-ph/0701272)

\bibitem[\protect\citeauthoryear{{Fragile}, {Mathews}, \& {Wilson}}{{Fragile}
  et~al.}{2001}]{fra01a}
{Fragile}, P.~C., {Mathews}, G.~J.,  \& {Wilson}, J.~R. 2001, \apj, 553, 955

\bibitem[\protect\citeauthoryear{{Fragile}, {Miller}, \&
  {Vandernoot}}{{Fragile} et~al.}{2005}]{fra05c}
{Fragile}, P.~C., {Miller}, W.~A.,  \& {Vandernoot}, E. 2005, \apj, 635, 157

\bibitem[\protect\citeauthoryear{{Gammie}, {McKinney}, \& {T{\' o}th}}{{Gammie}
  et~al.}{2003}]{gam03a}
{Gammie}, C.~F., {McKinney}, J.~C.,  \& {T{\' o}th}, G. 2003, \apj, 589, 444

\bibitem[\protect\citeauthoryear{{Hannikainen} et~al.}{{Hannikainen}
  et~al.}{2001}]{han01}
{Hannikainen}, D., {Campbell-Wilson}, D., {Hunstead}, R., {McIntyre}, V.,
  {Lovell}, J., {Reynolds}, J., {Tzioumis}, T.,  \& {Wu}, K. 2001, Astrophysics
  and Space Science Supplement, 276, 45

\bibitem[\protect\citeauthoryear{{Hawley}}{{Hawley}}{1991}]{haw91}
{Hawley}, J.~F. 1991, \apj, 381, 496

\bibitem[\protect\citeauthoryear{{Hawley}}{{Hawley}}{2000}]{haw00}
{Hawley}, J.~F. 2000, \apj, 528, 462

\bibitem[\protect\citeauthoryear{{Hawley}, {Smarr}, \& {Wilson}}{{Hawley}
  et~al.}{1984}]{haw84b}
{Hawley}, J.~F., {Smarr}, L.~L.,  \& {Wilson}, J.~R. 1984, \apjs, 55, 211

\bibitem[\protect\citeauthoryear{{Hughes}}{{Hughes}}{2001}]{hug01}
{Hughes}, S.~A. 2001, \prd, 64, 064004

\bibitem[\protect\citeauthoryear{{Koide}, {Shibata}, \& {Kudoh}}{{Koide}
  et~al.}{1999}]{koi99}
{Koide}, S., {Shibata}, K.,  \& {Kudoh}, T. 1999, \apj, 522, 727

\bibitem[\protect\citeauthoryear{{Komissarov}}{{Komissarov}}{2006}]{kom06}
{Komissarov}, S.~S. 2006, \mnras, 368, 993

\bibitem[\protect\citeauthoryear{{Kondratko}, {Greenhill}, \&
  {Moran}}{{Kondratko} et~al.}{2005}]{kon05}
{Kondratko}, P.~T., {Greenhill}, L.~J.,  \& {Moran}, J.~M. 2005, \apj, 618, 618

\bibitem[\protect\citeauthoryear{{Krolik}}{{Krolik}}{1999}]{kro99}
{Krolik}, J.~H. 1999, {Active Galactic Nuclei : From the Central Black Hole to
  the Galactic Environment} (Princeton, N.~J.~: Princeton University Press)

\bibitem[\protect\citeauthoryear{{Kumar} \& {Pringle}}{{Kumar} \&
  {Pringle}}{1985}]{kum85}
{Kumar}, S.,  \& {Pringle}, J.~E. 1985, \mnras, 213, 435

\bibitem[\protect\citeauthoryear{{Liu} \& {Melia}}{{Liu} \&
  {Melia}}{2002}]{liu02}
{Liu}, S.,  \& {Melia}, F. 2002, \apjl, 573, L23

\bibitem[\protect\citeauthoryear{{Lodato} \& {Pringle}}{{Lodato} \&
  {Pringle}}{2006}]{lod06}
{Lodato}, G.,  \& {Pringle}, J.~E. 2006, \mnras, 368, 1196

\bibitem[\protect\citeauthoryear{{Lubow}, {Ogilvie}, \& {Pringle}}{{Lubow}
  et~al.}{2002}]{lub02}
{Lubow}, S.~H., {Ogilvie}, G.~I.,  \& {Pringle}, J.~E. 2002, \mnras, 337, 706

\bibitem[\protect\citeauthoryear{{Maccarone}}{{Maccarone}}{2002}]{mac02}
{Maccarone}, T.~J. 2002, \mnras, 336, 1371

\bibitem[\protect\citeauthoryear{{Markovi{\'c} } \& {Lamb}}{{Markovi{\'c} } \&
  {Lamb}}{1998}]{mar98}
{Markovi{\'c} }, D.,  \& {Lamb}, F.~K. 1998, \apj, 507, 316

\bibitem[\protect\citeauthoryear{{McClintock} \& {Remillard}}{{McClintock} \&
  {Remillard}}{2005}]{mcc05}
{McClintock}, J.~E.,  \& {Remillard}, R.~A. 2005, in Compact Stellar X-ray
  Sources, in press (astro-ph/0306213)

\bibitem[\protect\citeauthoryear{{McKinney}}{{McKinney}}{2006}]{mck06}
{McKinney}, J.~C. 2006, \mnras, 368, 1561

\bibitem[\protect\citeauthoryear{{Miller} \& {Homan}}{{Miller} \&
  {Homan}}{2005}]{mil05}
{Miller}, J.~M.,  \& {Homan}, J. 2005, \apjl, 618, L107

\bibitem[\protect\citeauthoryear{{Nelson} \& {Papaloizou}}{{Nelson} \&
  {Papaloizou}}{2000}]{nel00}
{Nelson}, R.~P.,  \& {Papaloizou}, J.~C.~B. 2000, \mnras, 315, 570

\bibitem[\protect\citeauthoryear{Orosz \& Bailyn}{Orosz \&
  Bailyn}{1997}]{oro97}
Orosz, J.~A.,  \& Bailyn, C.~D. 1997, \apj, 477, 876

\bibitem[\protect\citeauthoryear{{Orosz} et~al.}{{Orosz} et~al.}{2002}]{oro02}
{Orosz}, J.~A., et~al. 2002, \apj, 568, 845

\bibitem[\protect\citeauthoryear{{Papadopoulos} \& {Font}}{{Papadopoulos} \&
  {Font}}{1998}]{pap98}
{Papadopoulos}, P.,  \& {Font}, J.~A. 1998, \prd, 58, 24005

\bibitem[\protect\citeauthoryear{{Papaloizou} \& {Lin}}{{Papaloizou} \&
  {Lin}}{1995}]{pap95a}
{Papaloizou}, J.~C.~B.,  \& {Lin}, D.~N.~C. 1995, \apj, 438, 841

\bibitem[\protect\citeauthoryear{{Rykoff} et~al.}{{Rykoff}
  et~al.}{2007}]{ryk07}
{Rykoff}, E.~S., {Miller}, J.~M., {Steeghs}, D.,  \& {Torres}, M.~A.~P. 2007,
  ArXiv Astrophysics e-prints

\bibitem[\protect\citeauthoryear{{Schnittman}, {Homan}, \&
  {Miller}}{{Schnittman} et~al.}{2006}]{sch06}
{Schnittman}, J.~D., {Homan}, J.,  \& {Miller}, J.~M. 2006, \apj, 642, 420

\bibitem[\protect\citeauthoryear{{Shakura} \& {Sunyaev}}{{Shakura} \&
  {Sunyaev}}{1973}]{sha73}
{Shakura}, N.~I.,  \& {Sunyaev}, R.~A. 1973, \aap, 24, 337

\bibitem[\protect\citeauthoryear{{Sobczak} et~al.}{{Sobczak}
  et~al.}{2000}]{sob00}
{Sobczak}, G.~J., {McClintock}, J.~E., {Remillard}, R.~A., {Cui}, W., {Levine},
  A.~M., {Morgan}, E.~H., {Orosz}, J.~A.,  \& {Bailyn}, C.~D. 2000, \apj, 531,
  537

\bibitem[\protect\citeauthoryear{{Stella}, {Vietri}, \& {Morsink}}{{Stella}
  et~al.}{1999}]{ste99}
{Stella}, L., {Vietri}, M.,  \& {Morsink}, S.~M. 1999, \apjl, 524, L63

\bibitem[\protect\citeauthoryear{{Wilson}}{{Wilson}}{1972}]{wil72}
{Wilson}, J.~R. 1972, \apj, 173, 431

\end{thebibliography}

\end{document}